\documentclass[12pt,letterpaper]{JHEP3}
\usepackage{amsmath,amssymb,amsfonts,xspace}
\usepackage{undertilde}

\newcommand{\url}[1]{\href{#1}  }

\def\sign{{\rm sign}}

\def\Re{\,{\rm Re}\, }

\def\({\left(}
\def\){\right)}
\def\[{\left[}
\def\]{\right]}

\newcommand{\de}{\mathrm{d}}

\newcommand{\I}{\mathrm{i}}
\newcommand{\e}{\mathrm{e}}

\newcommand{\p}{\partial}

\newcommand{\cC}{\mathcal{C}}
\newcommand{\cS}{\mathcal{S}}

\newcommand{\cK}{\mathcal{K}}
\newcommand{\cM}{\mathcal{M}}
\newcommand{\cN}{\mathcal{N}}
\newcommand{\cD}{\mathcal{D}}
\newcommand{\cE}{\mathcal{E}}

\newcommand{\cR}{\mathcal{R}}

\newcommand{\mbb}{\mathbb}
\newcommand{\mc}{\mathcal}
\newcommand{\f}{\frac}

\newcommand{\x}{{\bf x}}
\newcommand{\al}{\alpha}

\DeclareSymbolFont{AMSa}{U}{msa}{m}{n}
\DeclareSymbolFont{AMSb}{U}{msb}{m}{n}
\DeclareMathSymbol{\fieldR}{\mathalpha}{AMSb}{"52}

\newcommand{\N}{{\mathcal N}}
\newcommand{\kahler}{{K\"ahler}\xspace}
\newcommand{\hk}{{hyperk\"ahler}\xspace}
\newcommand{\qk}{{quaternion-K\"ahler}\xspace}

\newcommand{\cZ}{\mathcal{Z}}
\newcommand{\cO}{\mathcal{O}}

\renewcommand{\Re}{{\rm Re}}

\newcommand{\pa}{\partial}
\newcommand{\nn}{\nonumber}

\newcommand{\IR}{\mathbb{R}}
\newcommand{\IC}{\mathbb{C}}
\newcommand{\IZ}{\mathbb{Z}}

\newcommand{\IQ}{\mathbb{Q}}

\newcommand{\Tr}{\mbox{Tr}}

\newcommand{\sgn}{\mbox{sgn}}
\newcommand{\tzeta}{\tilde\zeta}

\def\rt {{\rm t}}

\def\bea{\begin{eqnarray}}
\def\eea{\end{eqnarray}}
\def\be{\begin{equation}}
\def\ee{\end{equation}}
\def\ba{\begin{align}}
\def\ea{\end{align}}
\def\bse{\begin{subequations}}
\def\ese{\end{subequations}}

\newcommand{\CA}{{\cal{A}}}

\newcommand{\CH}{{\cal{H}}}

\def\rf{r^{\flat}}
\def\vf{v^{\flat}}
\def\bz{{\bar z}}
\def\hu{{\hat u}}
\newcommand{\Li}{{\rm Li}}
\newcommand{\Nt}{\tilde{N}}

\newcommand{\SL}{\mathfrak{sl}}

\title{The automorphic NS5-brane}

\author{Boris Pioline
\\
{\it Laboratoire de Physique Th\'eorique et Hautes
Energies, \\ CNRS UMR 7589 and  
Universit\'e Pierre et Marie Curie - Paris 6,\\
4 place Jussieu, 75252 Paris cedex 05, France} \\
Email: \email{pioline@lpthe.jussieu.fr}
           }

\author{Daniel Persson\footnote{Also at \emph{Fundamental Physics, Chalmers University of Technology, SE-412 96, G\"oteborg, Sweden}.}\\
Physique Th\'eorique et Math\'ematique,\\ 
Universit\'e Libre de Bruxelles \& International Solvay Institutes,\\
ULB Campus Plaine C.P. 231, B-1050 Bruxelles, Belgium\\
Email: \email{dpersson@ulb.ac.be}}

\abstract{
Understanding the implications of $SL(2,\IZ)$ S-duality for the hypermultiplet moduli space 
of type II string theories has led to much progress recently 
in uncovering D-instanton contributions. In this work, 
we suggest that the extended duality group $SL(3,\IZ)$, which includes both 
S-duality and Ehlers symmetry, may determine the contributions of D5 and NS5-branes. 
We support this claim by automorphizing the perturbative
corrections to the ``extended universal hypermultiplet'', a five-dimensional universal 
$SO(3)\backslash SL(3,\IR)$ subspace which includes the string coupling,
overall volume, Ramond zero-form and six-form and NS axion. 
Using the non-Abelian Fourier expansion of the Eisenstein series attached
to the principal series of $SL(3,\IR)$, worked out by Vinogradov, Takhtajan 
and Bump many years ago, we extract the contributions of D(-1)-D5 and NS5-brane instantons,
corresponding to the Abelian and non-Abelian coefficients, respectively.
In particular, the  contributions  of $k$ NS5-branes 
can be summarized into a vector of wave functions $\Psi_{k,\ell}$, $\ell=0\dots k-1$,
as expected on general grounds.
We also point out that for more general models with a symmetric
moduli space $K\backslash G$, the minimal theta series of $G$ generates
an infinite series of exponential corrections of the form required for
``small" D(-1)-D1-D3-D5-NS5 instanton bound states. As a mathematical spin-off,
we make contact with earlier results in the literature about the spherical vectors 
for the principal series of $SL(3,\IR)$ and for minimal representations.}


\preprint{arXiv:0902.3274v4}

\begin{document}

\section{Introduction}

Understanding quantum corrections to hypermultiplet moduli spaces in type II Calabi-Yau compactifications is a long  and out-standing problem. One of the main challenges is to include the contributions from NS5-brane instantons which give rise to exponentially suppressed corrections to the moduli space metric of order $e^{-1/g_s^2}$ in the weak-coupling limit. In this work we propose that these NS5-brane instanton effects can be summed up in terms of a certain $SL(3,\mbb{Z})$-invariant Eisenstein series. To motivate this proposal, we begin by discussing some relevant aspects of hypermultiplet moduli spaces in type II Calabi-Yau compactifications, with particular emphasis on generalized mirror symmetry and dualities.

\subsection{Generalized mirror symmetry and \qk geometry}

Type II string theory and M-theory canonically associate 
two \qk (QK) spaces 
to any Calabi-Yau (CY) threefold $X$ \cite{Cecotti:1988qn},
\be
\label{mm}
\cM_{C}(X) \times \cM_{K}(X)\ ,
\ee
of real dimension $4h+4$ where
$h=h^{1,2}(X)$ and  $h=h^{1,1}(X)$, respectively. Both of these spaces 
$\cM=\cM_{C,K}(X)$ have a foliation by hypersurfaces $\phi=\text{const}$, such that near  $\phi=+\infty$, 
$\cM$ is topologically a fibration
\be
\label{fiber}
\tilde T \to \cM \to \IR_\phi \times \cK\ ,
\ee
where the $2h+3$ dimensional fiber $\tilde T$ is a circle bundle over a $2h+2$-dimensional
torus $T$, and  $\cK=\cK_{C,K}$ is the moduli space of complex structures (respectively, 
complexified \kahler structures) on $X$. Given any special \kahler metric on $\cK$,
the ``c-map'' construction \cite{Ferrara:1989ik}, or rather its ``quantum corrected version'' 
studied in \cite{Robles-Llana:2006ez,Alexandrov:2007ec}, 
produces a \qk metric on this fiber bundle, which agrees with the 
metric on $\cM$ in the ``weak coupling'' limit $\phi\to+\infty$, up to exponentially
suppressed corrections of order $\exp( -\cO( \e^{\phi/2}))$ and smaller. An outstanding question 
is to compute these corrections, which should encode interesting 
symplectic (resp. algebraic) invariants of $X$, and provide 
valuable information about the spectrum of type II string theories on $X$.

Indeed, the \qk spaces $\cM_{C,K}(X,Y)$ and associated special \kahler spaces $\cK_{C,K}$ 
appear as the moduli spaces of type IIA and type IIB string theories compactified 
on $X$ and $X\times S^1$ (respectively on $Y$ and $Y\times S^1$) as summarized 
in the table below  \cite{Cecotti:1988qn}:
%\begin{table}
\vspace*{3mm}
\be\nn
\begin{array}{|l|ccc|ccc|}\hline
    & &IIA/X& & &IIB/Y& \\ \hline
\IR^{3,1} & \cK_{K}(X) &\times& \cM_{C}(X) & \cK_{C}(Y) &\times& \cM_{K}(Y) \\
    & \downarrow  & & || &   \downarrow &  & || \\
\IR^{2,1} \times S^1  & \cM_{K}(X) &\times& \cM_{C}(X)  &  \cM_{C}(Y) &\times& \cM_{K}(Y)\\ \hline
\end{array}\\
\ee
%\end{table}
\vspace*{3mm}

\noindent This table calls for several important comments:
\begin{itemize}
\item[i)]The vertical arrows indicate the $c$-map relating the vector
multiplet (VM) moduli space $\cK$ in $D=4$ uncompact dimensions to the  
VM moduli space $\cM$ in $D=3$. In this case, the coordinate $\phi$
and the twisted torus $\tilde T$ in \eqref{fiber} correspond to the radius $e^{\phi/2}=R/l_P$
of the circle in 4D Planck units, the Wilson lines $\zeta^\Lambda, \tzeta_\Lambda$ of the  electric and 
magnetic vector fields in $D=4$ along $S^1$, and the NUT scalar $\sigma$ dual to the 
off-diagonal metric component  \cite{Ferrara:1989ik}. 
The circle $\tilde T/T=S_1$ parameterized by $\sigma$ 
has first Chern class $c_1=\de \zeta^\Lambda \wedge \de \tzeta_\Lambda + \chi \omega_\cK$,
where  $\omega_\cK$ is the \kahler class of $\cK$ and $\chi=\chi_X$ is the Euler number 
of $X$ in type IIA, or  $\chi=-\chi_Y$ 
in type IIB \cite{Robles-Llana:2006ez}. 

\item[ii)] The vertical equal signs indicate that the hypermultiplet (HM) moduli space
is identical in 3 and 4 dimensions; as a matter of fact, on the type IIA side $\cM_{C}(X)$
is also the HM moduli space in M-theory on $\IR^{4,1}\times X$, since the radius of the
M-theory circle is a vector multiplet. The coordinate $\phi$ 
and the twisted torus $\tilde T$ in \eqref{fiber} now correspond to the 
four-dimensional string coupling $e^{-\phi/2}$, the Wilson lines of the Ramond-Ramond
(RR) gauge fields on $H_{3}(X,\IZ)$ (respectively $H_{\rm even}(Y,\IZ)$), and the
Neveu-Schwarz (NS) axion.

\item[iii)] T-duality along the circle maps $IIA/X\times S^1$ to $IIB$ on the same  
CY three-fold $Y\equiv X$ times $S^1$ and exchanges the VM and HM moduli spaces in 
$D=3$ (in particular the radius $R/l_P$ is mapped to the string coupling $e^{\phi/2}$). This justifies 
the use of the same notation $\cM_{C,K}(\cdot)$ on the IIA and IIB sides.

\item[iv)] Mirror symmetry identifies $IIA/X\times S^1$ to $IIB/\tilde X \times S^1$ where
$Y\equiv \tilde X$ is the mirror CY three-fold to $X$ (in particular $h^{1,2}(X)=h^{1,1}(Y)$, 
$\chi_X=-\chi_Y$). At the level of the (2,2) SCFT it amounts to the well-supported identity
$\cK_C(X)=\cK_K(\tilde X)$, but at the ``second quantized'' level it requires
the more far-reaching identity \cite{Becker:1995kb}
\be
\label{secmir}
\cM_C(X)=\cM_K(\tilde X)\ ,\qquad \cM_K(X)=\cM_C(\tilde X)
\ee

\item[v)] The $c$-map construction mentioned above is accurate only in the limit $\phi\to +\infty$,
which corresponds to large radius  on the VM side, or small coupling on the HM side. 
$e^{-\cO(e^{\phi/2})}$ corrections away from this limit on the HM side correspond to 
D-brane instantons already in D=4  \cite{Becker:1995kb}, while on the VM side 
they correspond to Euclidean black holes in $\IR^3\times S^1$,
whose worldline winds around the circle. In either case,
D-instantons correspond mathematically 
to elements of the derived category $\mathcal{D}(X)$ of coherent sheaves for 
$\cM_K(X)$, or elements in the derived Fukaya category $\mathcal{F}(Y)$ of SLAG submanifolds
for $\cM_C(Y)$ (see e.g. \cite{Aspinwall:2004jr} for an introduction to these concepts).  
Thus, the equality \eqref{secmir} encompasses the
homological mirror symmetry conjecture \cite{MR1403918}. 

\item[vi)] In addition to the $e^{-\cO(e^{\phi/2})}$ D-instanton corrections mentioned in v),
one also expects $e^{-\cO(e^{\phi})}$ corrections, corresponding to Euclidean NS5-brane
wrapped on $X$ on the HM side, or to Kaluza-Klein (KK) monopoles (equivalently,
Taub-NUT gravitational instantons) with non-zero NUT charge along the circle.
These effects are predicted by the presumed growth of the D-instanton series \cite{Pioline:2009ia},
but the mathematical structure underlying them is far from clear at the moment.
They are the main subject of this note.

\item[vii)] Since $IIA/X\times S^1=M/X\times T^2$, the VM moduli space 
$\cM_K(X)$ must\footnote{Here,  we restrict to cases where,
unlike the situation in \cite{Aspinwall:1999ii}, $SL(2,\IZ)$ electric-magnetic duality is not 
broken to a finite index subgroup by
quantum corrections. We are grateful to N. Halmagyi for emphasizing this issue.\label{footbreak}}  
possess an isometric action of the modular group $SL(2,\IZ)$
of the torus $T^2$. Equivalently, S-duality of IIB string theory in 10 dimensions
implies that $\cM_K(Y)$ must have an (identical) isometric action of the modular 
group $SL(2,\IZ)$.

\end{itemize}
 
This last observation has been instrumental in the recent progress 
in understanding $\cM_K(X)$ \cite{RoblesLlana:2006is,Saueressig:2007dr,
RoblesLlana:2007ae,Saueressig:2007gi,Alexandrov:2008gh,Alexandrov:2009zh}: 
while the metric on $\cM_K(Y)$ in the limit of weak
coupling and large volume has a continuous isometric 
action  of $SL(2,\IR)$ \cite{Bodner:1990zm,Gunther:1998sc,Alexandrov:2008gh},
this isometric action is broken by the usual
worldsheet instanton corrections to $\cK_K(Y)$ and by a ``universal" one-loop 
correction \cite{Antoniadis:1997eg,Robles-Llana:2006ez,Gunther:1998sc}
proportional to $\chi_Y$. By restoring invariance under a 
discrete subgroup $SL(2,\IZ)$, it is possible to determine 
the corrections to the QK metric on $\cM_K(Y)$ due to D(-1) and D-1
instantons, i.e. to coherent sheaves with support on rational curves in $Y$ \cite{RoblesLlana:2006is}. 
Not surprisingly, these corrections are controlled by the same BPS
invariants which determine the tree-level worldsheet instanton corrections.
By mirror symmetry, these D(-1)-D1 instantons map to D2-brane instantons 
wrapping SLAG submanifolds $\gamma$ in $X$ whose homology class lies in a 
certain Lagrangian 
subspace of $H_3(X,\IZ)$ determined by the large complex structure limit. 
Using symplectic invariance, the effects of
D2-branes wrapping any homology class in $H_3(X,\IZ)$ were found in
\cite{Alexandrov:2008gh,Alexandrov:2009zh}, 
to linear order in perturbation around the weak coupling 
metric; by mirror symmetry this gives the instanton corrections to $\cM_K(Y)$ from arbitrary
D5-D3-D1-D(-1) instantons (or from D6-D4-D2-D0 black holes to 
$\cM_K(X)$), or mathematically from any element of the derived category 
$\cD(Y)$ (respectively, $\cD(X)$).

A key device in computing instanton corrections to the QK metric on $\cM=\cM_K(Y)$
is the Lebrun-Salamon theorem \cite{MR664330,MR1001707,Alexandrov:2008nk}, which relates
the QK metric on $\cM$ to the  complex contact structure on its twistor space $\cZ$.
As a consequence, the deformed geometry can be encoded in terms of complex contact transformations
between locally flat Darboux patches (which play a similar role as the holomorphic prepotential
for special \kahler spaces). The deformed metric can be obtained from the 
complex coordinates on $\mc{Z}$, 
also known as contact twistor lines, and from a complex valued (but non holomorphic) section
$e^{\Phi(x^{\mu}, z)}$ of $H^0(\cM,\cO(2))$, known as the contact potential, which
determines the \kahler potential on $\cZ$.

Using these twistorial techniques, the QK metric on $\cM_K(Y)$ including quantum
corrections from all D-instantons was obtained to linear order
 in \cite{Alexandrov:2008gh, Alexandrov:2009zh}. It involves
invariants $n_\gamma$, presumed to be equal
to the generalized Donaldson-Thomas invariants introduced in \cite{MR2302500,ks}.
An essentially identical structure has emerged in the study of instanton corrections
to the (\hk) moduli space of $\cN=2$ Seiberg-Witten theories on $\IR^{1,2}\times S^1$
in \cite{Gaiotto:2008cd}, and in fact directly inspired the construction 
in \cite{Alexandrov:2008gh}.

These developments have left out 
the outstanding problem of computing the subleading $e^{-\cO(e^{\phi})}$ corrections 
to $\cM_{C,K}$ from NS5-brane instantons wrapped on $X$ (or equivalently 
from KK-monopoles with non-zero NUT charge on $S^1$). While the 
$SL(2,\IZ)$ symmetry which proved so powerful in determining the D-instanton corrections
could in principle be used to convert the D5-instanton corrections into NS5-branes
(or D6-branes into KK-monopoles), it is not immediately clear 
how to covariantize these contributions under the complicated 
$SL(2,\IZ)$ action found in  \cite{Alexandrov:2008gh}.

\subsection{Uncovering $SL(3,\mbb{Z})$, and the extended universal hypermultiplet}

In this note, we employ a different strategy, and investigate how 
invariance under a larger discrete group, $SL(3,\IZ)$,  may constrain 
the NS5-brane contributions. This discrete symmetry is most easily
seen in M-theory on $\IR^{1,2}\times T^2\times X$: indeed, after
reduction on a torus $T^{d}$ of any dimension $d$ and dualization of the Kaluza-Klein
connection into scalars, the Einstein-Hilbert Lagrangian in 
$D=d+3$ dimensions leads to a $SL(d+1,\IR)$ invariant non-linear
sigma model in $D=3$ dimensions: this $SL(d+1,\IR)$ symmetry
includes the manifest $SL(d,\IR)$ symmetry from 
diffeomorphism invariance on $T^d$, and a multiplet of $d$ (non-commuting) 
$SL(2,\IR)$ Ehlers symmetries \cite{Ehlers:1957zz} apparent in the two-step reduction 
$D=d+3\to D=4 \to D=3$. The discrete subgroup $SL(d,\IZ)$ of global diffeomorphisms
of the torus should clearly remain a symmetry of the quantum theory, but it is reasonable to
assume that a larger discrete subgroup $SL(d+1,\IZ)$ is in fact unbroken
 quantum mechanically\footnote{The fact
that the discrete Ehlers symmetry  is unbroken quantum mechanically 
in M-theory on $\IR^{1,2}\times T^8$ follows by intertwining the geometric $SL(8,\IZ)$
and  T-duality $SO(7,7,\IZ)$ symmetries, see e.g. \cite{Obers:1998fb}; 
in the heterotic string on $T^6$, Ehlers symmetry
is related to S-duality by a sequence of T-dualities \cite{Bakas:1994ba}. We are
not aware of a similar derivation in the $\cN=2$ setting.}.
In the case at hand with $d=2$, we postulate
that quantum corrections preserve a $SL(3,\IZ)$ subgroup of $SL(3,\IR)$,
larger than the $SL(2,\IZ)$ S-duality warranted by diffeomorphism invariance.

At this stage we should  warn the reader against a possible confusion
with another $SL(3,\IZ)$ symmetry expected by duality to the heterotic
$E_8\times E_8$ on $\IR^{1,2}\times T^3\times K3$: indeed, when (and only when) 
$X$ admits a K3-fibration with a section, heterotic-type II duality \cite{Kachru:1995wm}
predicts that 
\be
\label{hetii}
\cM(K3) = \cM_C(X)\ ,\qquad \cM(T^3)=\cM_K(X)\ ,
\ee
where $\cM(K3)$ 
parametrizes the Ricci-flat metric and $E_8\times E_8$ bundle on $K3$, while
$\cM(T^3)$ parametrizes the flat metric and $E_8\times E_8$ bundle on $T^3$
and the scalars dual to the Kaluza-Klein connections and the $U(1)^{16}$
Abelian gauge fields in three dimensions. Just as on the M-theory side, 
$\cM(T^3)$ has an obvious $SL(3,\IZ)$ symmetry, enhanced to $SL(4,\IZ)$ by 
Ehlers-type transformations. The M-theory $SL(3,\IZ)$ action on $\cM_K(X)$ is part of
the heterotic $SL(4,\IZ)$ action on $\cM(T^3)$, but intersects the geometric
$SL(3,\IZ)$ action only along the $SL(2,\IZ)$ S-duality subgroup.

Before discussing how a discrete $SL(3,\IZ)$ symmetry can be preserved by 
quantum corrections, we must understand how the continuous symmetry
group $SL(3,\IR)$ acts on the weak coupling, large volume limit of $\cM_K(X)$.
We shall argue that in this limit, the moduli space $\cM_K(X)$ decomposes as a
product
\be
\label{prosl3}
\cM_K(X) \sim \frac{SL(3,\IR)}{SO(3)} \times \cR_K(X) \times \IR^{3h_{1,1}(X)}\ ,
\ee
where $\cR_K(X)$ is a space of real dimension $h_{1,1}(X)-1$, which
appears as the VM moduli space in M-theory on $\IR^{4,1}\times X$.
$SL(3,\IR)$ acts on the first factor in \eqref{prosl3} by the usual non-linear action, leaves
the second factor inert and acts linearly on $(\IR^{3})^{\otimes h_{1,1}(X)} $. 

In particular, we claim that the hypermultiplet moduli space $\cM_K(Y)$ 
in type IIB string theory compactified on $Y$ admits a universal sector
$\cM_u= SO(3)\backslash SL(3,\IR)$, of real dimension 5,
which consists of the ten-dimensional axio-dilaton $\tau$, 
the overall volume $V=t^3$ of $Y$ in string units, 
the Wilson line $c_0$ of the RR six-form potential on $Y$, and the 
four-dimensional NS  axion $\psi$. Despite the fact that this universal sector
does not carry any QK metric, we refer to it as the ``extended universal hypermultiplet",
to distinguish it from the ``universal hypermultiplet'' \cite{Strominger:1997eb,Becker:1999pb,
Antoniadis:1997eg,Antoniadis:2003sw,Anguelova:2004sj,Alexandrov:2006hx}, 
which has real dimension 4 and carries, at tree-level, a $SU(2,1)$ 
invariant QK metric. The latter is universal in the sense that it appears as 
a subfactor in  any ``c-map'' construction \cite{Ferrara:1989ik}. However,  
it is unclear whether a finite covolume discrete subgroup of $SU(2,1)$ 
should stay unbroken in general. However, see \cite{Bao2009} for a discussion
of this possibility when $X$ is a rigid CY threefold. 

Finally, let us mention that our identification of $SL(3,\IZ)$ as the unbroken discrete
subgroup of $SL(3,\IR)$ is tentative: it is quite possible that only a finite index subgroup
of $SL(3,\IZ)$ may be unbroken, as it happens with $SL(2,\IZ)$ electric-magnetic duality 
on the vector multiplet side. It is rather easy to adapt our considerations to this case, and
it may in fact be the key to resolve a shortcoming of our proposal to be discussed presently.

\subsection{Summing up NS5-brane instantons}

Having postulated that $SL(3,\IZ)$ is preserved a the quantum level, we shall demonstrate that 
this symmetry potentially determines a subset of the
NS5-brane corrections, once the tree-level worldsheet instantons and 
the one-loop  correction are given.
The adverb ``potentially'' is in order since our specific proposal \eqref{conj} 
leads to unexpected terms which blow up at weak coupling (see \eqref{ep} below). 
%Nevertheless,  we are hopeful that these terms can be interpreted physically,
%or removed with a minor change of our set up, and 
%that our main conclusions will continue to hold.
Our approach is very close in spirit to the one taken in \cite{Kiritsis:1997em}, where the $SL(3,\IZ)$ U-duality symmetry of type II string on $\IR^{7,1}\times T^2$ was used
to determine the contributions of $(p,q)$ strings to $R^4$ couplings in the effective action.
Technically, however, we require the more sophisticated automorphic forms of $SL(3,\IZ)$
constructed in \cite{Pioline:2004xq} in the context of BPS membranes.

As mentioned above, quantum corrections to the QK moduli space 
$\mc{M}_K(Y)$ are conveniently encoded in complex coordinates on its twistor space $\mc{Z}$,
together with the contact potential $e^{\Phi(x^{\mu}, z)}$. Taking the conjectured $SL(3,\mbb{Z})$-invariance at face value, we shall propose a non-perturbative 
completion of the contact potential $e^{\Phi (x^{\mu}, z(x^\mu))}$ restricted to a certain section
$z(x^\mu)$ of  $\mc{Z}$, in terms of a certain non-holomorphic Eisenstein series $E(g; s_1, s_2)$ attached to the principal continuous series of $SL(3,\mbb{R})$. Relying
on the thirty-year old analysis of this Eisenstein series  by Vinogradov and 
Takhtajan \cite{MR527787} and Bump \cite{MR765698}, 
we show that for the special values $(s_1,s_2)=(3/2, -3/2)$ the Fourier expansion of $E(g; s_1, s_2)$ reproduces the correct universal\footnote{i.e., depending only on  the generalized universal 
hypermultiplet moduli, and on the Euler number of $Y$.}
tree-level and one-loop corrections to the hypermultiplet metric. 
Moreover, the non-Abelian Fourier expansion of $E(g; s_1, s_2)$  predicts 
an infinite series of exponentially suppressed contributions 
at weak coupling, of two distinct types:
\begin{enumerate}
\item[  i)] the Abelian contributions, 
of order $e^{-S_{p,q}}$ where $S_{p,q}$ is independent of the NS-axion, given in \eqref{InstantonActionDD} below, 
can be interpreted as contributions from bound states of 
$p$ D5 and $q$ D$(-1)$-instantons. In particular, the instanton action $S_{p,q}$ 
correctly reproduces the mass formula for D0-D6 branes 
on the type IIA side \cite{Dhar:1998ip,Rasheed:1995zv,Larsen:1999pp}.
Via the $c$-map, the summation measure \eqref{InstantonMeasureDD} should be related to
the D0-D6 bound state degeneracies
predicted by the Mac Mahon function \cite{Denef:2007vg}, but checking this
lies beyond the scope of this work.
\item[ii)] the non-Abelian contributions, of order $e^{-S_{Q,p,k}}$ given
in \eqref{genS} below, have a non-trivial dependence on the NS-axion $\psi$,
and can be interpreted as instanton corrections from bound states of $Q$ 
D(-1)-instantons and $(p,k)$ 5-branes. Their action \eqref{genS} 
follows from the D5-D(-1) action \eqref{InstantonActionDD} by S-duality,
after subtracting a moduli independent contribution $e^{- 2\pi \I q d \alpha/k}$ in 
\eqref{vtransformed}. The latter is responsible for the apparent divergence of \eqref{genS} 
at $k=0$. The summation measure \eqref{NonAbelianMeasure} is obtained
from  \eqref{InstantonMeasureDD} by replacing $p\to d, q\to Q/d^2$ where
$d=\gcd(p,k)$ and multiplying by the phase $e^{2\pi \I Q \alpha/(d k)}$.
In representation theoretic terms, as explained in Appendix D, 
this provides the exact real and adelic spherical vectors for the principal
continuous series of $SL(3,\mbb{R})$ beyond the semi-classical limit obtained in 
\cite{Pioline:2004xq}.
\end{enumerate}
A general property of the non-Abelian Fourier coefficients, and therefore
of the NS5-brane instantons, is that they satisfy a wave function property:
namely, the non-Abelian Fourier expansion can be carried out for 
different choices of polarization, e.g. \eqref{nonab} or 
\eqref{nonab2}, and the corresponding summands $\Psi_{k,\ell}$
and $\tilde\Psi_{k,\ell'}$  are related by Fourier transform, Eq.
\eqref{changepol} below. It is tempting to conjecture that the wave function
$\Psi_{1,0}$ describing the contribution of one NS5-brane is related
to the topological string amplitude, possibly along the lines of 
\cite{Dijkgraaf:2002ac,Kapustin:2004jm}. 

While our main emphasis is on the universal sector, we also
speculate on the $SL(3,\IZ)$-invariant completion of the ``non-universal" contributions, which
include D3 and D1-instantons, and suggest that  in the context of ``magic" supergravity models
with a symmetric hypermultiplet moduli space \eqref{MJ} 
the minimal theta series associated to ${\rm QConf}(J,\IZ)$ may resum the contributions of
``very small instantons", i.e. those whose charges satisfy $I_4=\pa I_4 = \pa^2 I_4=0$,
where $I_4$ is the quartic invariant for the duality group ${\rm Conf}(J)$. In particular,
\eqref{Sming} should provide the general action for bound states of $(p,k)$-5 branes, 
$N^a$ D3-branes, $\tilde N_a$ D1-branes and $\tilde Q$ D-instantons, at least when 
the D1 and D(-1) instanton charges are induced from the D3 brane charge via 
\eqref{indcharge}. Thus, we for the first time provide a physical interpretation of the spherical vector 
$f_K$ for the minimal representation of any quasiconformal group ${\rm QConf}(J,\IZ)$, which has been
known for simply laced groups in the split real form since \cite{Kazhdan:2001nx}. From this point of view,
the puzzling cubic phase appearing in $f_K$ simply originates from the D-instanton axionic coupling
by an $SL(2,\IZ)$ transformation, after subtracting out a moduli-independent contribution 
as in ii) above.

We now mention some limitations of our proposal. Firstly, in order to 
obtained the deformed QK metric the contact
potential $e^{\Phi(x^{\mu}, z)}$ should be supplemented by the twistor lines. It would be very interesting to understand how to incorporate $SL(3,\mbb{Z})$-invariance in this 
context\footnote{The  $SL(2,\mbb{Z})$ action on the instanton-corrected twistor lines has recently been
clarified, and is in fact identical to the tree-level action after suitable field redefinitions \cite{Alexandrov:2009qq}.}. 
In addition, our proposal predicts puzzling 
perturbative contributions beyond the expected tree-level and one-loop terms,
which grow like negative genus contributions or diverge faster than linearly at large volume.
It is conceivable that these terms could be avoided by postulating invariance under 
a finite index subgroup of $SL(3,\IZ)$, or may be attributed to hitherto unknown physical effects.
Moreover, our proposal for the $SL(3,\IZ)$-invariant completion of  non-universal effects is 
tentative only, and would require a better understanding of the $SL(3,\IR)$ action 
on the non-universal sector of the hypermultiplet moduli space.

\subsection{Outline}

The rest of this article is organized as follows. In Section \ref{univ}, we discuss the geometry
 of the extended universal hypermultiplet, and work out the decomposition 
 \eqref{prosl3} in the one-modulus case.  In Section 3, we review
how the $SL(2,\IZ)$ symmetry of the HM moduli
 space $\cM_K(X)$ can be restored after including suitable D(-1) and D1- 
 instanton contributions,
 and show how $SL(3,\IZ)$ may similarly be restored by including 
 NS5-brane contributions (together with D5 and D(-1)-instantons). 
 Moreover, we identify the NS5-brane
 contributions as certain non-Abelian Fourier coefficients of the corresponding
 automorphic form, and comment on their wave function property. 
 In Appendix A we collect some results on the constant terms of
minimal and generalized Eisenstein series with respect to certain parabolic subgroups.
In Appendix B we give a detailed derivation of the non-Abelian Fourier 
 expansion of the minimal Eisenstein series for $SL(3,\IZ)$.
In Appendix C a certain key integral is computed in the saddle point approximation.
Finally, in Appendix D, we give a representation theoretic viewpoint on 
non-Abelian Fourier expansions, and extract the exact spherical vector 
for the  principal series of $SL(3,\mbb{R})$.
 \newpage

\section{The extended universal hypermultiplet \label{univ}}

In this section, we show that the symmetric space $\cM_u=SO(3)\backslash SL(3,\IR)$
can be viewed as a universal sector of the HM space  $\cM_K(X)$ in the large
volume, weak coupling limit, and work out the decomposition 
 \eqref{prosl3} in the one-modulus case.

\subsection{$SO(3)\backslash SL(3,\IR)$ as a hypermultiplet moduli space}

The five-dimensional symmetric space $\cM_u$ may be parametrized
in the Iwasawa gauge by the coset 
\be
\begin{split}
g =& \left( {\nu^{-1/6}}\sqrt{\tau_2} \right)^{H_p} \cdot
 \left( \nu^{-1/3} \right)^{H_q}\cdot  
e^{\tau_1 E_p}\cdot e^{c_0 E_q}\cdot e^{\psi E}\\
=&\begin{pmatrix} {\nu^{1/6}}/\sqrt{\tau_2}  & & \\ & {\nu^{1/6}} \sqrt{\tau_2}   & \\ && \nu^{-1/3} \end{pmatrix}
\cdot  \begin{pmatrix}1&\tau_1 &\psi + \tau_1 c_0  \\ & 1 & c_0   \\ && 1 \end{pmatrix}\ ,
\label{iwa1}
\end{split}
\ee
where $E_i=\{E,E_p,E_q,H_p,H_q,F_p,F_q,F\}$ form a basis of the Lie algebra of
$\SL(3,\IR)$, such that any linear combination $\sum \utilde{E}_i E_i$ with $\utilde{E}_i\in \IR$
is represented in the triplet representation by
\be
\begin{pmatrix} -\utilde{H}_p & \utilde{E}_p & \utilde{E} \\
-\utilde{F}_p & -\utilde{H}_q+\utilde{H}_p & \utilde{E}_q \\
-\utilde{F} & -\utilde{F}_q & \utilde{H}_q 
\end{pmatrix} \ ,
\ee
The maximal compact subgroup $K=SO(3)$ is generated by antisymmetric matrices,
i.e. by $E_p+F_p, E_q+F_q, E+F$.

The right-invariant metric on $\cM_u$ is obtained from the right-invariant form
$\theta=\de g\cdot g^{-1}$ projected along $K$ via
\be
\label{ds1}
\de s^2 = \frac12\Tr[(\theta+\theta^\rt)^2]=\frac{\de\nu^2}{3\nu^2} + \frac{\de\tau_1^2+\de\tau_2^2}{
\tau_2^2} +\nu \, \frac{(\de\psi+\tau_1 \de c_0)^2 +\tau_2^2 \de c_0^2}{\tau_2} \ .
\ee
The Killing vectors generating the right-action of $SL(3,\IR)$ on $\cM_u$ are given by 
\be
\label{kilsl3}
\begin{split}
E&=\pa_{\psi}\ ,\qquad
E_p =\pa_{\tau_1}-c_0 \pa_{\psi} \ ,\qquad 
E_q=\pa_{c_0}\ ,\\
H_p &= 2 \tau_1 \pa_{\tau_1} 
+ 2   \tau_2 \pa_{\tau_2} +\psi \pa_{\psi}  -c_0 \pa_{c_0}  \ ,\\
H_q &=2 c_0 \pa_{c_0}-3  \nu \pa_{\nu} +  \psi \pa_{\psi}
  -{\tau_1} \pa_{\tau_1} -{\tau_2} \pa_{\tau_2} \ ,\\
F_p &=  
   -\psi \pa_{c_0}  -2 {\tau_1} {\tau_2} \pa_{\tau_2} +\left(\tau_2^2-\tau_1^2\right)\pa_{\tau_1} \ ,\\
F_q& = c_0^2 \pa_{c_0}
   -c_0 (3 \nu \pa_{\nu}  - \psi \pa_{\psi} + {\tau_1} \pa_{\tau_1}
  + {\tau_2} \pa_{\tau_2}) -\psi\pa_{\tau_1} 
   -(\nu\tau_2)^{-1}(\pa_{c_0}-{\tau_1}\pa_{\psi} )\ ,\\
F& =
 \psi (\psi \pa_{\psi}+  c_0 \pa_{c_0}  -3 \nu  \pa_{\nu} 
   +\tau_1 \pa_{\tau_1} +\tau_2 \pa_{\tau_2} ) 
   +c_0  [ (\tau_1^2 -\tau_2^2 ) \pa_{\tau_1}
   +2   \tau_1 \tau_2 \pa_{\tau_2} ]\\
  & -(\nu\tau_2)^{-1} [ (\tau_1^2+\tau_2^2)\pa_\psi -\tau_1 \pa_{c_0} ]
\end{split}
\ee
For later reference, we record the Laplace-Beltrami operator on $\cM_u$,
equal to the quadratic Casimir of the $\SL(3,\IR)$ action \eqref{kilsl3},
\be
\label{lapl}
\cC_2 = \tau_2^2 (\pa_{\tau_1}^2+  \pa_{\tau_2}^2 ) + 3 \pa_\nu ( \nu^2\pa_\nu) 
+ \frac{1}{\nu\tau_2} ( \pa_{c_0} - \tau \pa_{\psi} )( \pa_{c_0} - \bar\tau \pa_{\psi} )\ .
\ee
There is also an invariant differential operator $\cC_3$ of third order in derivatives,
corresponding to the cubic Casimir given in \eqref{c2c3}. 

The parametrization \eqref{iwa1} was chosen such the $SL(2,\IR)$ subgroup
corresponding to matrices of the form
\be
\label{sl2tau}
\begin{pmatrix}\  a &\ b & \ \\ c & d & \ \\ \ & \ & 1\ \end{pmatrix}\ ,\quad ad-bc=1\ ,
\ee
acts by fractional linear transformations on $\tau\equiv\tau_1+\I\tau_2$
and linearly on $(c_0,\psi)$,
\be\label{SL2Z}
\tau \mapsto \frac{a\, \tau +b}{c\, \tau + d} \, , \qquad
\begin{pmatrix} c_0 \\ \psi \end{pmatrix} \mapsto
\begin{pmatrix} d & -c \\ -b & a  \end{pmatrix}
\begin{pmatrix} c_0 \\ \psi \end{pmatrix}\ ,
\qquad \nu \mapsto \nu\ .
\ee

To recognize \eqref{ds1} as a hypermultiplet moduli space metric, let us change
variables to $(\phi,t,\zeta,\tzeta,\sigma)$ defined by
\be
\label{chvar}
%V= t^3\ ,\quad 
\nu=\frac{e^{-3\phi/2}}{t^{3/2}}\ ,\quad
\tau_2 = \frac{e^{\phi/2}}{t^{3/2}}\ , 
\quad \tau_1 = \zeta\ , \quad
c_0 = \tzeta\ ,\quad \psi = -\frac12 ( \sigma + \zeta \tzeta ) 
\ee
The metric \eqref{ds1} becomes
\be
\label{dsuniv}
\de s^2 =  \de \phi^2 + 3 \frac{\de t^2}{t^2} + e^{-\phi} 
\left( t^{-3} \de\tzeta^2 + t^3 \de\zeta^2   \right) + \frac14 e^{-2\phi} 
\left( \de\sigma + \tzeta \de\zeta - \zeta \de\tzeta \right)^2 \ .
\ee
This is the standard $c$-map metric associated to a special \kahler
manifold $\cK(X)$ with cubic prepotential $F=-\frac 16 \kappa_{abc} z^a z^b z^c$ \cite{Ferrara:1989ik},
restricted to the locus $z^a=\I t \, r^a, \zeta^a=\tzeta_a=0$, where 
$r^a$ is a fixed reference value for the \kahler modulus $t^a$.
Comparing \eqref{chvar} to \cite{Alexandrov:2008gh}, we can identify
$\tau$ as the 10D type IIB axio-dilaton, $c_0=-\int_Y A^{(6)} + \dots$ 
as the Ramond-Ramond six-form background, $\psi$ as the 4D Neveu-Schwarz
axion, $t^3\equiv V$ as the volume of $Y$ in string units. The four-dimensional string coupling is then 
\be
g_4\equiv \f{1}{\tau_2 \sqrt{8V} } = \frac{1}{\sqrt8} e^{-\phi/2}
\label{4dStringCoupling}
\ee
(the factor of $\sqrt8$ is conventional), while the Heisenberg Killing vectors act
as 
\be
\label{Heissym}
E_p= \pa_\zeta + \tzeta \pa_\sigma\ ,\qquad E_q= \pa_{\tzeta} - \zeta \pa_\sigma\ ,\qquad
E_k = -2 \pa_\sigma\ .
\ee

It is perhaps worth noting that although a further restriction to the locus $t=1$ produces 
the $SU(2,1)$-invariant metric on the universal hypermultiplet, 
$SU(2,1)$ does not act on the five-dimensional manifold \eqref{dsuniv}.
The stabilizer of the locus $t=\text{const}$ is the semi-direct product of $\IR^+$
generated by $H_p+H_q$ and  the Heisenberg group $N$:
 \be
\label{heisper}
\begin{pmatrix}\ 1& \ m\  &  -p+\frac12 m n\ \\  & 1 & n   \\  & & 1 \end{pmatrix}\ :\quad 
(\zeta,\tzeta,\sigma) \mapsto (\zeta+m,\,\tzeta+n,\,
\sigma+2p- n \zeta+ m \tzeta)\ .
\ee
On the other hand, the stabilizer of the locus $t^2+e^{-\phi}  \tzeta^2/t=\text{const}$
is the semi-direct product of $\IR^{2}$ generated by $E_p,F_q$ and 
the $SL(2,\IR)$ subgroup
\be
\label{sl2s}
\begin{pmatrix} A &\  &\ B\ \\  & 1 &  \\ C\ &\  &\ D \end{pmatrix}\ ,\quad AD-BC=1\ ,
\ee
which acts as 
\be
S\mapsto \frac{A\,S+B}{C\,S+D}\ ,\qquad 
\begin{pmatrix} x  \\ y \end{pmatrix} \mapsto
\begin{pmatrix} D & -C \\ -B & A  \end{pmatrix}
\begin{pmatrix} x \\ y \end{pmatrix}\ ,
\qquad z \mapsto z\ ,
\ee
where
\be
\begin{split}
S&\equiv \frac12(-\sigma+\zeta\tzeta) +\I\, e^{\phi} \sqrt{1+\frac{e^{-\phi}}{t^3} \tzeta^2}\ ,\qquad
z \equiv t^2+\frac{e^{-\phi}}{t} \tzeta^2\ ,\qquad\\
&\qquad \begin{pmatrix} x  \\ y \end{pmatrix}  \equiv 
\frac{1}{e^\phi t^3 + \tzeta^2}
\begin{pmatrix} \tzeta  \\ \zeta(e^\phi t^3 + \tzeta^2) + \frac12 \tzeta(\sigma-\zeta\tzeta)
 \end{pmatrix} 
 \end{split}
\ee
This $SL(2,\IR)$ subgroup is just the Ehlers symmetry alluded to in the introduction, written in 
a somewhat unusual field basis. Note that for $\tzeta=0$, the complex variable $S$ reduces to the four-dimensional axio-dilaton,
%\be
$S\big|_{\tzeta=0}=-\f{1}{2}\sigma+\I e^{\phi}$.
%\ee 

In the sequel, we shall assume that physical amplitudes are invariant under 
$SL(3,\IZ)$, the group of integer valued, determinant one matrices. In particular,
this includes the $SL(2,\IZ)$ subgroup \eqref{sl2tau} with $a,b,c,d$ integer,
 the $SL(2,\IZ)$ subgroup \eqref{sl2s} with $A,B,C,D$ integer, and the
Heisenberg group \eqref{heisper} with $m,n, -p+\frac12 mn$ integer.

\subsection{The universal sector in the one-modulus case}

To clarify in what sense $SO(3)\backslash SL(3,\IR)$ is a universal sector of $\cM_K$, 
we now study the decomposition \eqref{prosl3} in the one-modulus case, 
when $\cM_{K}$ is the \qk manifold  $\cM= SO(4) \backslash G_{2(2)}$, 
obtained by the $c$-map procedure from a special \kahler manifold
$\cK_K$ with prepotential $F=-(X^1)^3/X^0$. The geometry of 
this symmetric space was studied in detail 
in  \cite{Gunaydin:2007qq}, whose notations we adhere to\footnote{Except for
the following changes of notation: $\tau\to z,\zeta^\Lambda\to \zeta^\Lambda/\sqrt2,  
\tzeta_\Lambda\to -\tzeta_\Lambda/\sqrt2, \sigma\to-\frac12\sigma$.}. 
The \qk metric
\be
ds^2 = 2 \left(
u \,\bar u +
v \, \bar v +
e^1 \, \bar e^1 +
E_1 \, \bar E_1 \right)\ ,
\ee
with 
\be
\begin{split}
u =& \frac{e^{-\phi/2}}{ 4\, t^{3/2}}
\left( -\de\tzeta_0 -  z \de\tzeta_1
+ 3  z^2 \de\zeta^1 -  z^3 \de\zeta^0 \right)\\
v =& \frac12 \de\phi - \frac{i}{4} e^{-\phi} (
\de\sigma - \zeta^0 \de\tzeta_0 -\zeta^1 \de\tzeta_1 
+  \tzeta_0 \de\zeta^0 +\tzeta_1 \de\zeta^1)\\
e^1 =&  \frac{i \sqrt{3}}{2t} \de z  \\
E_1 =&
-\frac{e^{-\phi/2}}{  4\sqrt{3}\,t^{3/2}}
\left( -3 \de\tzeta_0 - \de\tzeta_1 (\bar z+2 z)
+3  z (2\bar z+ z) \de\zeta^1 -3 \bar z  \,z^2 \de\zeta^0 \right)
\end{split}
\ee
has a $G_{2(2)}$ isometric action, and therefore a $SL(3,\IR)\subset G_{2(2)}$ isometric action.
This action corresponds to right multiplication on the coset representative in the
Iwasawa gauge (here $z\equiv b+\I t$),
\be
\label{iwag2} 
e =
t^{-Y_0} \cdot e^{\sqrt{2} b Y_+} \cdot e^{-\frac12 \phi H} \cdot e^{-\frac{1}{\sqrt2}\zeta^0 {E_{q_0}} -\frac{1}{\sqrt2} \tzeta_0
{E_{p^0}}} \cdot e^{-\sqrt{\frac{3}{2}} \zeta^1 {E_{q_1}} -\frac{1}{\sqrt{6}} \tzeta_1 {E_{p^1}}} \cdot
e^{-\frac12\sigma E} \ ,
\ee 
followed by a compensating $SO(4)$ left-action. 

The $SL(3,\IR)$ subgroup of $G_{2(2)}$ is generated 
by the longest roots with respect to a split Cartan torus. 
A system of coordinates adapted to the $SL(3)$ action is obtained 
by choosing instead a coset representative in the (non-Iwasawa) gauge
\be
\label{noniwag2}
e=\nu^{\frac{Y_0}{6}+\frac{H}{4}}\cdot 
\tau_2^{\frac{Y_0}{2}-\frac{H}{4}}\cdot 
e^{-\frac{1}{\sqrt2} \tau_1 E_{q_0}}\cdot
e^{-\frac{1}{\sqrt2}c_0 E_{p^0}}\cdot
e^{\psi E_k}\cdot e^{\sqrt{\frac{3}{2}} u_1 E_{q_1} -  \sqrt2 u_2 Y_+ + \sqrt{\frac{3}{2}} u_3 F_{p^1}}
\ee
In this way, the coordinates $(\nu,\tau_2,\tau_1,c_0,\psi)$ parametrize $\cM_u$
as in \eqref{iwa1}, with the same transformations \eqref{kilsl3} as before, while the 
real coordinates $(u_1,u_2,u_3)$ transform 
linearly in the triplet representation of $SL(3)$. In these variables, the metric can be written as
\be
\de s^2_{\cM}= \de s^2_{\cM_u}
+ \de\vec u^2 + \left( 1 + \frac13 \vec u^2 \right) 
(\vec u \wedge \de \vec u)^2 +  \CA^{ijk} u_i u_j  \de u_k 
\ee
where the contractions of the three-vectors $\vec u, \de\vec u$ and $\vec u \wedge \de \vec u$
are performed with the $3\times 3$ symmetric matrices $M, M, M^{-1}$, respectively, where 
$M=g^\rt g$. Here $\CA_{(ij)k}$ are $SL(3)$ invariant forms on $\cM_u$. The origin
of the various terms can be understood by writing it schematically as follows:
\be
\de s^2_{\cM}= (\de s_{\cM_u} + \CA \, u\, u\, \de u)^2
+ \de\vec u^2 +  (\vec u \wedge \de \vec u)^2
\ee
reflecting the decomposition $14=8+3+\bar 3=5+3+3+3$ of $\mathfrak{g}_2$ under
$\mathfrak{so}(3) \subset\SL(3)$.
 Note that there is no translational symmetry along the $\vec u$ variables:
indeed the triplet of generators $(E_{q_1}, -\sqrt6 Y_+,F_{p^1})$ differs from
$\pa/\pa u_i$ at linear order in $u_j$. 

The relation between the two sets of coordinates can be found  by 
determining the $SO(4)$ left action needed to cast \eqref{noniwag2} in 
Iwasawa form \eqref{iwag2}. We suppress the details and quote
only the result:
\be
\label{sl3var}
\begin{split}
  \phi &= -\frac{1}{2} \log \left(\frac{\nu \Delta^2}{\tau_{2}}\right) \ ,\qquad
 t= \frac{\Delta^{1/2}}
   {\nu^{1/6} \tau_2^{1/2}
   \left(\hu_{3}^4+\left(\hu_{1}^2+\hu_{2}^2+2\right)
   \hu_{3}^2+1\right)}\ ,\\
 b&= -\frac{\hu_{2}+\hu_{1} \hu_{3}}{\nu^{1/6}
   \sqrt{\tau_{2}} \left(\hu_{3}^4+\left(\hu_{1}^2+\hu_{2}^2+2\right)
   \hu_{3}^2+1\right)},\\
 \zeta^0&= \tau_{1}-\frac{\tau_{2} \hu_{3}}{\Delta}
   \left(\hu_{2} \hu_{3} \hu_{1}^3-\hu_{1}^2+\hu_{2} \hu_{3}
   \left(\hu_{2}^2+\hu_{3}^2+3\right)
   \hu_{1}+\hu_{2}^2\right)\ ,\\
\zeta^1&=
   \frac{\sqrt{\tau_{2}}}{\nu^{1/6} \Delta}\ \left(\hu_{2} \hu_{3}
   \hu_{1}^2-\left(\hu_{3}^2+1\right) \hu_{1}+\hu_{2} \hu_{3}
   \left(\hu_{2}^2+\hu_{3}^2+1\right)\right)\ ,\\
\tzeta_0&= c_0 - \frac{\Xi}{(\nu \tau_2)^{1/2} \Delta}
 \ ,\\
\tzeta_1&= -\frac{3}
   {\nu^{1/3} \Delta} \left(\hu_{3} \left(\hu_{2}^2+\hu_{3}^2\right) \hu_{1}^2-\hu_{2}
   \hu_{1}+\hu_{3} \left(\hu_{2}^2+\hu_{3}^2+1\right)^2\right)\ ,\\
\sigma &= -2
   \psi - \tau_{1} c_0 + \frac{ \tau_{2} \hu_{3} c_0}{\Delta}
   \left(\hu_{2}
   \hu_{3} \hu_{1}^3-\hu_{1}^2+\hu_{2} \hu_{3}
   \left(\hu_{2}^2+\hu_{3}^2+3\right)
   \hu_{1}+\hu_{2}^2\right) 
   -\frac{\tau_{1} \Xi-\tau_2 \Xi'}{  \Delta\sqrt{\nu  \tau_{2}}} \ ,
\end{split}
\ee
where
\be
\begin{split}
\hu_1&= \frac{\nu^{1/6}}{\sqrt{\tau_2}} \left(u_1+\psi  u_3 +\tau_1
   (u_2 +c_0 u_3 )\right)\ ,\quad
\hu_2 = \nu^{1/6} \sqrt{\tau_2}
    (u_2+c_0  u_3)\ ,\quad
\hu_3 \nu^{-1/3} u_3\ ,\\
\Delta&=\hu_{3}^6+\left(2 \hu_{2}^2+3\right)
   \hu_{3}^4+\left(\hu_{2}^4+3 \hu_{2}^2+3\right) \hu_{3}^2+\hu_{1}^2
   \left(\hu_{2}^2+\hu_{3}^2\right) \hu_{3}^2-2 \hu_{1} \hu_{2}
   \hu_{3}+1\ ,\\
 \Xi&=  \hu_{2} \hu_{3}
   \left(\hu_{2}^2+\hu_{3}^2\right)
   \hu_{1}^2-\left(\hu_{2}^2-\hu_{3}^2\right) \hu_{1}+\hu_{2} \hu_{3}
   \left(\hu_{2}^4+\left(2 \hu_{3}^2+3\right) \hu_{2}^2+\hu_{3}^4+3
   \hu_{3}^2+3\right)\ ,\\
 \Xi'&= \left(\hu_{3} \left(\hu_{2}^2+2 \hu_{3}^2\right) \hu_{1}^3-\hu_{2}
   \hu_{1}^2+\hu_{3} \left(\hu_{2}^4+3 \left(\hu_{3}^2+1\right)
   \hu_{2}^2+2 \hu_{3}^4+6 \hu_{3}^2+3\right) \hu_{1}+\hu_{2}
   \hu_{3}^2\right)\ .  
   \end{split}
\ee

A similar decomposition holds for any \qk space $\cM$
given by the $c$-map of a special \kahler manifold $\cK$ with cubic prepotential
\be
F = -\frac16 \kappa_{abc} X^a X^b X^c / X^0\ ,
\ee
where $\kappa_{abc}$ is the norm form of a Jordan algebra $J$
of degree three. In this case, $\cM$ is 
a symmetric space \cite{Gunaydin:1983bi,Gunaydin:2004md,Gunaydin:2005zz,Pioline:2006ni}
\be
\label{MJ}
\cM=[SU(2)\times \widetilde{\rm Conf}(J)]\backslash {\rm QConf}(J)\ ,
\ee
where ${\rm QConf}(J)$ and $\widetilde{\rm Conf}(J)$ are the quasi-conformal and 
compact conformal groups associated to $J$. The root lattice of  ${\rm QConf}(J)$ 
admits a two-dimensional projection to the root lattice of $G_{2(2)}$, with
a non-trivial multiplicity $h$ for the short roots, and with the zero weights corresponding
to the 5-dimensional duality group ${\rm Str}_0(J)$ together with 
the non-compact Cartan generators of $SL(3)$.  Using a suitable (non-Iwasawa)
gauge, the right-invariant metric can be written  as  
\be
%\begin{split}
\de s_{\cM}%&
= \de s^2_{\cM_u} +    \de s_{\cR}^2 + (\de\vec u^a)^2 +
\left( 1 + \frac13 (\vec u^a)^2 \right) 
\left( \frac16 \kappa_{abc}  \vec u^b \wedge \de \vec u^c\right)^2 
%&
+ \frac16 \kappa_{abc} \, \CA^{ijk} \, u_i^a u_j^b  \de u_k^c  \ ,
%\end{split} 
\ee
where $a=1\dots h$, where $\cR={\rm Aut}(J)\backslash {\rm Str}_0(J)$ is the 
vector multiplet space in 5 dimensions, given by the cubic hypersurface \cite{Gunaydin:1983bi},
\be
\frac16\kappa_{abc} r^a r^b r^c=1\ ,
\ee
of real dimension $h-1$.
The coordinate $t$ on $\cM_u$ is then the overall scale of the \kahler classes,
$t^a = t\, r^a$, while the coordinates $u_i^a$ are related to the RR Wilson lines
by a generalization of \eqref{sl3var},  e.g. to leading order in $u_3^a$,
\be
\begin{split}
b^a&=-u_2^a+\dots\ ,\qquad   \qquad  \quad \quad\ \ 
 \tzeta_0= c_0+ \frac16 \kappa_{abc}  u_{2}^a u_2^b (u_{1}^c+\tau_1\, u_{2}^c) + \dots,\quad\\
 \zeta^a &= -  (u_{1}^a+\tau_1\, u_{2}^a)+\dots,\qquad
 \tzeta_a=  \frac12 \kappa_{abc} [ u_{2}^b (u_{1}^c +\tau_1\, u_{2}^c) +  
 e^\phi t^b u_3^c ]+ \dots
\end{split}
\ee
These formulae should remain correct in the large volume, weak coupling limit,
even when the intersection form $\kappa_{abc}$ is not the norm form of a 
Jordan algebra, and $\cM_K$ not a symmetric space. 

\section{$SL(3,\IZ)$ Eisenstein series and NS5-instantons}

While $\cM_K(X)$ admits an isometric action of $SL(3,\IR)$  in the strict 
weak coupling, large volume limit, quantum corrections to the metric generically 
break all continuous isometries. In this section, we show
that a  discrete subgroup $SL(3,\IZ)$ may be restored, provided 
that quantum corrections take a suitable form.

\subsection{Quaternionic-\kahler geometry and contact potential}

The \qk metric on $\cM$ is conveniently encoded in the \hk potential, 
a $SU(2)$-invariant, degree one homogeneous function $\chi$ on the Swann bundle 
$\cS$, which provides a \kahler potential for the \hk 
metric on $\cS$  in all complex structures \cite{MR1096180,deWit:2001dj}. 
Here $\cS$ is a 
$\IR^4/\IZ_2$ bundle over $\cM$ (equivalently a $\IC^\times$ bundle over the twistor space $\cZ$ of $\mc{M}$), 
which carries a canonical \hk metric
with an isometric $SU(2)$ action and homothetic Killing 
vector $\kappa$ \cite{MR1096180}.
Thus, one may choose coordinates $x^\mu$ on $\cM$ and
$(\vf,\bar \vf, z,\bz)$ on $\IR^4/\IZ_2$ such 
that \cite{Neitzke:2007ke,Alexandrov:2008nk,Alexandrov:2008gh},
\be
\label{defchi}
\chi = 4|\vf| \frac{(1+z\bar z)}{|z|} \, e^{\Re[\Phi(x^\mu,z)]}\ .
\ee
where $\Phi(x^\mu,z)$, a complex function holomorphic in $z$, is known as the
 ``contact potential''. In order to extract the metric on $\cS$ or on $\cM$, the \hk 
potential $\chi$  should also be supplemented by the ``twistor lines'', i.e. by 
a set of holomorphic functions $u^i(x^\mu,z)$ on $\cZ$ such that $(\vf, u^i)$ 
provides a set of local complex coordinates on $\cS$. 
Importantly, any isometry
of $\cM$ can be combined with a suitable action on $(\vf,\bar \vf, z,\bz)$ to produce a tri-holomorphic isometry of $\cS$, leaving $\chi$ invariant.  In the presence of one continuous isometry,
$\Phi(x^\mu,z)$ can be taken to be independent of $z$, but this is not possible in general.
However, using the $SU(2)$ action, it is in principle possible to recover $\Phi(x^\mu,z)$
for any $z$ from the knowledge of its restriction to any section $z(x^\mu)$. 
In this note, we shall restrict our attention to this ``restricted" contact potential $\Phi$
for a suitable section $z(x^\mu)$,  leaving for future work the 
determination of the twistor lines and of the contact potential 
away from the section  $z(x^\mu)$.

In type IIB string theory compactified on $Y$,  the contact potential on the HM moduli space
$\cM_K(Y)$,  including the effects of the tree-level $(\alpha')^3$ correction,
tree-level world-sheet instantons and one-loop correction was determined 
in \cite{RoblesLlana:2006is,Alexandrov:2008nk,Alexandrov:2008gh}:
\be
\label{phiclas}
\begin{split}
e^{\Phi_{\rm pert}} =& \frac{\tau_2^2}{2} \,V
- \frac{\sqrt{\tau_2}}{16(2\pi)^3}\,\chi_Y
\left[ 2\zeta(3)\,\tau_2^{3/2}+\frac{2\pi^2}{3}\,\tau_2^{-1/2}
\right]\\&+
\frac{\tau_2^2}{4(2\pi)^3}\sum\limits_{k_a> 0} n_{k_a}^{(0)}\,
\Re\left[ \Li_3 \left( e^{2\pi \I k_a z^a} \right) + 2\pi k_a t^a\,
\Li_2 \left( e^{2\pi \I k_a z^a} \right)  \right]  \, ,
\end{split}
\ee
where $V\equiv\frac16 \kappa_{abc}t^a t^b t^c=t^3$ is the volume of $Y$,
$n_{k_a}^{(0)}$ is the BPS invariant in the homology class 
$k_a \gamma^a\in H_2(Y,\IZ)$, $\Li_s(x)=\sum_{n=1}^{\infty} x^n/n^s$ is the polylogarithm
and $\zeta(s)$ is Riemann's zeta function. 
%We remind the reader that in terms of the four-dimensional string coupling (\ref{4dStringCoupling}), the classical contribution to $e^{\Phi_{\rm pert}}$ is just the expected tree-level term $g_s^{-2}/2$. 

In  the weak coupling, large volume limit, only the first term in \eqref{phiclas} remains.
The \hk potential $\chi$ is then invariant under the $SL(2,\IR)$ 
groups \eqref{sl2tau} and \eqref{sl2s}, respectively, provided the prefactor 
$\rf\equiv \vf| {(1+z\bar z)}/{|z|}$ in \eqref{defchi} transforms as
\be
\rf \mapsto   \rf\, |c\, \tau+d| \ ,\qquad 
\rf \mapsto \rf\, |C\, S+D|^2 
\sqrt{\frac{1+ z^{3/2} x^2 S_2}{1+\frac{z^{3/2}(D x-B y)^2 S_2}{  |C\, S+D|^2 } }} \ , 
\ee
respectively. This invariance is spoiled, however, when the 
other terms in \eqref{phiclas} are included. Of course, $\chi$ could
always be made invariant by adjusting the transformation rule of 
$\rf$, but this will in general not lead to a tri-holomorphic action. For this reason,
we do not allow any deformation of the $SL(3,\IR)$ action on the 
coordinates $x^\mu$ and $\rf$\footnote{The consistency of this
assumption in the case of $SL(2,\IZ)$ has been checked recently in \cite{Alexandrov:2009qq}.}.  
Instead, we allow for deformations of the contact potential $\Phi(x^\mu)\equiv 
\Phi(x^\mu,z(x^\mu))$, but
assume  that there exists a choice of section $z(x^\mu)$
such that  $\Phi(x^\mu,z(x^\mu))$ retains its tree-level transformation property.

In \cite{RoblesLlana:2006is}, it was shown that an $SL(2,\IZ)$ subgroup 
of  \eqref{sl2tau} could be
restored by adding to the perturbative potential \eqref{phiclas} a suitable 
combination of D-instantons and $(m,n)$-string instantons, 
\be
e^{\Phi_{\rm inv}} = \frac{\tau_2^2}{2} \,V
+\frac{\sqrt{\tau_2}}{8(2\pi)^3}
\!\!\sum_{k_a\gamma^a\in H_2^+(Y)\cup\{0\}}\!\!\!\!\!\,
 n_{k_a}^{(0)}\,
{\sum\limits_{m,n}}'\frac{\tau_2^{3/2}}{|m\tau+n|^3}\(1+2\pi |m\tau+n|k_a t^a\)\, e^{-2\pi S_{m,n, k_a}}\, ,
\label{phiinv}
\ee
where $n_{0}^{(0)}=-\chi_Y/2$, 
\be
S_{m,n,k_a} = k_a | m \tau+n |\, t^a-\I k_a (m c^a +n  b^a)
\ee
and the primed sum runs over pairs of integers $(m,n)\neq (0,0)$. Thus,
$SL(2,\IZ)$ invariance is powerful enough to determine these types of 
instanton corrections, which with  our current understanding of string theory could 
not be computed from first principles. As a strong consistency check, 
it was shown that \eqref{phiinv} reproduces the expected behavior near 
the conifold \cite{Saueressig:2007dr}. 

\subsection{Automorphizing under $SL(3,\IZ)$}
\label{automorphizing}

Our aim is to show that similarly, invariance under a discrete subgroup 
$SL(3,\IZ)$ of $SL(3,\IR)$ can be restored by including NS5-brane and D5-brane contributions.  
For simplicity we concentrate  on the $(\alpha')^3$ and $g_s$ ``universal" corrections  
in the first line of \eqref{phiclas}, which depend only on the extended universal sector
$(\nu,\tau_2,\tau_1,c_0,\psi)$ and on the Euler number $\chi_Y$. Factoring out the tree-level contribution, we require that
\be
\label{phipert3}
e^{\Phi} = \frac{\tau_2^2 \,V}{2} \left( 1 
+ E(g) \right) \ ,\qquad
\ee
where $E(g)$ is  an $SL(3,\IZ)$-invariant function such that, at weak coupling, 
\be
\label{epert}
E(g) = - \frac{\chi_Y}{8(2\pi)^3} \left(  2\zeta(3) \, V^{-1} +2 \frac{\pi^2}{3}\, V^{-1} \tau_2^{-2}  + \dots 
\right)
\ee

While our knowledge of automorphic forms of $SL(3,\IZ)$ is rather limited,
some general principles and a few explicit examples are well understood.
As explained e.g. in \cite{Pioline:2003bk}, $G(\IZ)$-invariant functions
on $K\backslash G(\IR)$ can be constructed from 
\begin{itemize}
\item[i)] a unitary representation
$\rho$ of $G(\IR)$ in an (infinite dimensional) Hilbert space $\CH$, 
\item[ii)] a ``spherical'' K-invariant vector\footnote{The term 
``spherical'' requires both K-invariance and suitable decrease at infinity.
If $\rho$ does not admit a spherical vector, $f_K$ can be replaced by a vector 
in the lowest K-type, but \eqref{overl} then leads to a section of a non-trivial homogeneous vector
bundle over $K\backslash G(\IR)$.} $f_K\in \CH$, and 
\item[iii)] a $G(\IZ)$-invariant 
distribution $f_\IZ$ on $\CH$.
\end{itemize} 
Moreover, $f_\IZ$ can often be obtained adelically
from spherical vectors $f_p$ of the representation $\rho$ over the $p$-adic
number field $\IQ_p$ for all primes $p$. With these ingredients, one may write
\be
\label{overl}
E(g) =  \langle f_\IZ | \rho(g^{-1}) | f_K \rangle\ ,
\ee
which is well defined on $K\backslash G(\IR)/G(\IZ)$, due to the 
$G(\IZ)$- and $K$-invariance of $f_\IZ$ and $f_K$, respectively.
Moreover, if $\rho$ is an irreducible representation, such that 
any $G$-invariant operator $\cO$ acts on $\CH$ as a scalar, then 
$E(g)$ is an eigenmode of $\cO$ with the same eigenvalue, now acting
as a $G$-invariant differential operator on $K\backslash G(\IR)$.
In particular, the eigenvalue of $E(g)$ under the Laplace-Beltrami
operator \eqref{lapl} on $K\backslash G(\IR)$ is equal to the value of the 
quadratic Casimir $\cC_2$ in the representation $\rho$.

In the context of $R^4$ couplings in 8 dimensions \cite{Kiritsis:1997em}, an elementary example of a 
$SL(3,\IZ)$-invariant function was constructed (see also \cite{Obers:1999um} for a physics discussion 
of this type of Eisenstein series): 
\be
\label{eismin}
E(g;s) = \sum_{m\in\IZ^3\backslash\{(0,0,0)\}} {\Big(m^\rt  M m\Big)^{-s}}
\ee
where $M=g^\rt g$. The sum converges absolutely
when $\Re(s)>3/2$, and can be analytically continued in the rest of the $s$-plane, except
for a pole at $s=3/2$. This is the automorphic form associated to the 
representation $\rho$ on homogeneous functions of degree $-2s$
in three variables -- also known as the minimal representation.
The spherical vector is just $f_K=(m^\rt m)^{-s}$, and
$f_\IZ$ is the Dirac distribution on the lattice $\IZ^3$ minus the 
origin. The values of the quadratic and cubic Casimirs are given by 
\be
\label{casmin}
\cC_2=\frac23 s(2s-3)\ ,\qquad \cC_3=-\frac{2}{27}s(2s-3)(4s-3)\ ,
\ee
satisfying the relation
\be
\label{c3c2rel}
4\, \cC_2^3 - 27\, \cC_3^2 +3\, \cC_2^2 =0\ .
\ee
The infinitesimal generators of the minimal representation 
are spelled out in Appendix \ref{ap_min}.  

More generally, one may consider  the principal Eisenstein series 
\cite{MR0176097, MR0579181,MR0486315,MR527787,MR765698} 
(see also \cite{Pioline:2003bk,Pioline:2004xq} for a physics discussion)
\be
\label{eisgen}
E(g;s_1,s_2) = 
\frac{1}{4\zeta(2s_1)\zeta(2s_2)} \sum_{m,n}
%_{
%\begin{array}{c}
%\ss\{ (m,n)\in \Zint^3\otimes \Zint^3 \ :\\
%\ss m\neq 0,\, n\sim n+z m \neq 0\}
%\end{array}}\!\!\!
{\Big(m^\rt  M m\Big)^{-s_1}}\  
{\Big( n^\rt   M^{-1} n \Big)^{-s_2} }\, ,
\ee
where the sum runs over pairs of  non-zero integer vectors $\vec m,\vec n$ such that
$m^\rt n=0$. The automorphic form (\ref{eisgen})
is attached to the principal continuous representation obtained by induction
from the minimal parabolic (or ``Borel") subgroup
\be
\label{defpmin}
P_{\rm min}= \left\{\begin{pmatrix}a_3 &*&*\\ &a_2&*\\ &&a_1\end{pmatrix}\Big|\ a_1 a_2 a_3=1\right\}
\ee 
via the character\footnote{More generally, one could multiply $\chi$ by
$\prod_{i=1\dots 3}[\sign(a_i)]^{\epsilon_i}$, and obtain in this way the supplementary
continuous series; however the signs $\epsilon_i$ must be chosen to be $+$ in order
for the $SL(3,\IZ)$ Eisenstein series not to vanish. The supplementary series may become
relevant if only a finite index subgroup of $SL(3,\IZ)$ was unbroken.
We are grateful to S. Miller for consultations about this point.} 
\be
\chi_{s_1,s_2}(p)=|a_1|^{\frac23(s_1+2s_2)}\, |a_2|^{\frac23(s_1-s_2)}\, |a_3|^{-\frac23(2s_1+s_2)}\ .
\label{charind}
\ee
In contrast to the minimal representation, the principal series has independent 
quadratic and cubic Casimirs, 
\be
\label{casgen}
\cC_2=\frac43(s_1^2+s_2^2+s_1 s_2)-2(s_1+s_2)\ ,\qquad
\cC_3=-\frac{2}{27}(s_1-s_2)(2s_2+4 s_1-3)(2s_1 + 4 s_2-3)\ .
\ee
The infinitesimal generators of the continuous principal representation 
are spelled out in Appendix \ref{ap_prin}. The sum in (\ref{eisgen}) converges absolutely for 
\be\Re(s_1)>1\ ,\qquad \Re(s_2)>1\ ,
\label{convdom}
\ee
and may be meromorphically continued to other values of $(s_1, s_2)$  \cite{MR0579181,MR0419366,MR527787}. Singularities
arise at the six lines in the $(s_1, s_2)$  plane where \eqref{c3c2rel} is obeyed, namely
\be
\label{linmin}
s_1=0\ ,\qquad s_1=1 \ ,\qquad s_2=0\ ,\qquad s_2=1 \ ,\qquad 
s_1+s_2=\frac12\ ,\qquad s_1+s_2=\frac32\ ,
\ee
where $E(g;s_1,s_2)$ becomes proportional\footnote{This can be seen by comparing
the constant terms, see Appendix \ref{ConstantTerms}. The matching  
of the Abelian and non-Abelian Fourier coefficients appears less obvious but should 
hold on general grounds.  
Note that none of the lines in \eqref{linmin} lies in the convergence domain \eqref{convdom}.} to the minimal Eisenstein series $E(g;s)$ \eqref{eismin}.

Finally, one may also consider representations induced from
the maximal parabolic subgroup  
\be
\label{defpmax}
P_{\rm max}= \left\{\begin{pmatrix}*&*&*\\{}*&*&*\\&&*\end{pmatrix}\right\}\ ,
\ee
and construct
\be
\label{eisgenphi}
E(g;s_1,s_2,\phi) = 
\sum
{\Big(m^\rt  M m\Big)^{-s_1-2s_2}}\, 
\phi\left( \tau_{m,n}, \bar\tau_{m,n} \right) \, ,
\ee
where
\be
\tau_{m,n} =
\frac{m^\rt   M n +\I \sqrt{(m^\rt   M m)\  (n^\rt   M n)- (m^\rt   M n)^2}}
{m^\rt   M m}\ ,
\ee
$\phi(\tau,\bar\tau)$ is a non-holomorphic modular form of $SL(2,\IZ)$ with weight $-2s_2$, 
\be
\phi\left( \frac{a\,\tau+b}{c\,\tau+d}, \frac{a\,\bar\tau+b}{c\,\bar\tau+d}\right)
=|c\,\tau+d|^{2s_2} \, \phi(\tau,\bar\tau)\ ,
\ee
and the sum runs over pairs of non-zero integer vectors $m,n$, modded out by the
equivalence relation $n\sim n+m$.
For $\phi=\tau_2^{-2s_2}$, this reproduces \eqref{eisgen} up to a normalization factor. 
According to \cite{MR0222215}, these induced representations
exhaust all irreducible unitary representations of $SL(3,\IR)$.

In the case at hand, the values of the quadratic and cubic Casimirs 
can be easily determined by acting with the invariant differential operators
$\cC_2$ and $\cC_3$ in \eqref{lapl} on the two perturbative terms in  \eqref{epert}:
\be
\label{c23}
\cC_2=3\ ,\qquad \cC_3=0\ .
\ee
This determines  $(s_1,s_2)$ in \eqref{eisgen} to be one of the six 
values (which lie away from the singular locus \eqref{c3c2rel})
\be
\left\{ \left(\frac32, \frac32\right), \left(\frac32, -\frac32\right), \left(-\frac32, \frac32\right), 
\left(-\frac12,-\frac12\right), \left(-\frac12,\frac52\right),
\left(\frac52,-\frac12\right) \right\}\ .
\ee
The Weyl group of $SL(3)$ acts by permuting these values, and the
resulting Eisenstein series \eqref{eisgen} are identical, up to an overall
$(s_1,s_2)$-dependent constant. There is no loss of generality in choosing
$(s_1,s_2)=(3/2,-3/2)$. Thus, we tentatively propose that the $SL(3,\IZ)$
invariant function appearing in \eqref{phipert3} is given by 
\be
\label{conj}
E(g) = - \frac{\chi_Y}{4(2\pi)^3} \,\zeta(3)\, E(g; 3/2, -3/2)\ ,
\ee
where the generalized Eisenstein series on right-hand side is given by \eqref{eisgen} 
with\footnote{Different choices of $(s_1,s_2)$ simply amount to permuting  
the various terms in  \eqref{esl3sl2}  and \eqref{ep}.}
$(s_1,s_2)=(3/2,-3/2)$. 
%We emphasize that the Eisenstein series $E(g;s_1,s_2)$ (\ref{eisgen}) is regular for these particular values of $(s_1,s_2)$, in marked contrast to the minimal Eisenstein 
%series $E(g;s)$ (\ref{eismin}) which has a pole at $s=3/2$. 

More generally, it is natural to conjecture
that the worldsheet instanton sum in \eqref{phiinv} is subsumed into a 
sum of $SL(3,\IZ)$-invariant functions
\be
\begin{split}
\tilde E(g) &=
\frac{\zeta(3)}{2(2\pi)^3} \, 
\!\!\sum_{k_a\gamma^a\in H_2^+(Y)\cup\{0\}}\!\!\!\!\!\,
 n_{k_a}^{(0)}\\
&\qquad \sum_{m,n} \left( \frac{m^\rt   M m\  n^\rt   M n- (m^\rt   M n)^2}{m^\rt  M m}\right)^{3/2}\,
\left(1+2\pi k_a r^a \sqrt{m^\rt M\,m} \right) \, e^{-2\pi S_{\vec m, k_a}}  
\label{phiinvsl3}
\end{split}
\ee
where $n_0^{(0)}=-\chi_Y/2$,
\be
S_{\vec m,k_a} =k_a r^a \sqrt{m^\rt M\, m} +\I k_a \,\vec u^{\,a} \, \vec m \ ,
\ee
and the second sum runs over the same set of integers as in \eqref{eisgen}.
In the sequel we shall restrict ourselves to the universal contributions \eqref{conj} 
corresponding to $k_a=0$, leaving
a study of \eqref{phiinvsl3} to future work.  

We should stress that we do not know whether \eqref{eisgen} 
is the only automorphic form of $SL(3,\IZ)$ with the infinitesimal parameters \eqref{c23}. 
It seems reasonable however to take it as a working assumption, and see what kind of
quantum corrections it predicts.

\subsection{Perturbative and D-instanton contributions}
\label{Sec:ConstantTerms}

In order to justify our proposal \eqref{conj}, we should check  that the perturbative terms
in \eqref{epert}, and indeed the whole D-instanton series \eqref{phiinv} predicted on the
basis of $SL(2,\IZ)$ duality, are reproduced in the large volume  limit $\nu\to0$. 
The mathematical prescription is to extract the ``constant term'' with respect to the 
maximal parabolic subgroup \eqref{defpmax},
i.e. the zero-th Fourier coefficient with respect to $(c_0,\psi)$,
\be
E_{P_{\rm max}}(\nu,\tau_2,\tau_1)\equiv \int_0^{1} \de c_0\, \int_0^{1}  \de \psi \, 
E\left(\nu,\tau_1,\tau_2,c_0,\psi;\frac32,-\frac32\right)\ .
\ee
Since $(c_0,\psi)$ transforms as a doublet under \eqref{sl2tau}, the result must be
invariant under $SL(2,\IZ)\subset P_{\rm max}$. Using the general results due 
to Langlands \cite{MR0579181,MR0419366}, or the explicit computation 
in  \cite{MR527787}, we find (see \eqref{p2} in Appendix A)
\be
\label{esl3sl2}
\begin{split}
E_{P_{\rm max}}(\nu,\tau_2,\tau_1) &= \tau_2^{-3/2} \,V^{-1}\, E_{3/2}(\tau)
  -1080 \,  \zeta (3) \,\zeta '(-4) \,\tau_2^{9/2} V^3 \, E_{-1/2} (\tau)\\
  &\qquad +120 \, \zeta (3)  \,\tau_2^{3/2} V \, E_{-3/2}(\tau)
  \end{split}
\ee
where $E_s(\tau)$ is the standard non-holomorphic $SL(2,\IZ)$ Eisenstein series
\be
\label{sl2eis}
E_s(\tau) = \sum_{(m,n)\neq (0,0)}
\left(
\frac{\tau_2}{|m+n\tau|^2}
\right)^s \ ,
\ee
whose Fourier decomposition is given by the Chowla-Selberg formula 
(see e.g. \cite{Green:1997tv, Obers:1999um} for a physicis discussion)
\be
\label{sl2exp}
\begin{split}
E_s(\tau) &= 2\zeta(2s)\, \tau_2^s + 2\sqrt{\pi}\, \tau_2^{1-s}\,
\frac{\Gamma(s-1/2)}{\Gamma(s)} \zeta(2s-1)  \\
& \qquad\qquad +  \frac{2 \pi ^s \tau_2^s}{(2\pi)^{1/2-s}\Gamma(s)}
\sum_{\tilde m_1\ne 0} \sum_{m_2\ne 0}
\left| \tilde m_1 \right|^{2s-1}
\cK_{s-1/2} \left(2\pi \tau_2 |\tilde m_1 m_2|\right)
e^{2\pi i \tilde m_1  m_2 \tau_1}\ .
\end{split}
\ee
To simplify formulas, we have defined the rescaled (modified) Bessel function:
\be
\label{asK}
{\cal K}_t(x)\equiv x^{-t}K_t( x)
= \sqrt{\f{\pi}{2}} x^{-(t+1/2)} e^{-x} \Big( 1 + \cO( 1/x) \Big)\,.
\ee
The first term $E_{3/2}(\tau)$ in \eqref{esl3sl2} indeed reproduces \eqref{phiinv} 
with $k_a=0$, and therefore the
two perturbative terms in \eqref{epert}. 

To analyze the remaining terms in \eqref{esl3sl2},
it is useful to go to the weak coupling limit $\tau_2\to\infty$, where only the first
line of \eqref{sl2exp} contributes for each of the Eisenstein series $E_s$ appearing
in \eqref{esl3sl2}. In effect, this amounts to extracting the constant
term 
\be
E_{P_{\rm min}}(\nu,\tau_2) = \int_0^1 \de \tau_1 E_{P_{\rm max}}(\nu,\tau_2,\tau_1)
\label{Sec:ConstantTermMinimalParabolic}
\ee
with respect to the minimal parabolic (or Borel) subgroup \eqref{defpmin} (see Eq. \eqref{eisct}): 
\be
\label{ep}
\begin{split}
E_{P_{\rm min}}(\nu,\tau_2)  &= 2\zeta(3) \, V^{-1} + \frac{2 \pi ^2}{3} \,\tau_2^{-2} \, V^{-1}
+ \frac{405}{\pi ^6}\, \zeta (3)^2 \,\zeta (5)\, V   \\
&\qquad + 180 \, \zeta (3) \,\zeta'(-4) \, \tau_2^4 \,V^3 
 + 180 \, \zeta (3) \, \zeta'(-4)  \, \tau_2^4   \,V \, + 2  \zeta (3) \,\tau_2^6 \, V^3\ .
 \end{split}
\ee
Multiplying out by the prefactor $\tau_2^2 V/2$ from \eqref{phipert3},  we see that
the terms on the second line behave like perturbative contributions with negative
genus $-2$ and $-3$, while the last term on the first line behaves like a tree
level contribution which grows up like the square of the volume. These are 
the puzzling terms mentioned in the introduction. It could
be that such divergent terms arise perturbatively
(in analogy with the $\log g_s$  term encountered in $R^4$ couplings \cite{Kiritsis:1997em}),
or that the proposal \eqref{conj} is too naive. Nevertheless,
it is instructive to analyze the implications of our proposal at finite volume and 
coupling, in the hope that these issues can be resolved in the future with minor changes to our set-up.

\subsection{Non-Abelian Fourier expansion and the minimal Eisenstein series\label{secnonab}}

At finite volume and coupling, terms with non trivial dependence on $c_0$
and $\psi$ will start contributing, corresponding in the type IIB context to
D5 and NS5-brane instantons (or, in the IIA context, to D6 and KK-monopoles winding
along $S^1$). However, due to the non-Abelian nature of
the Heisenberg group $N$ in \eqref{heisper}, it is not possible to diagonalize
translations in $\tau_1, c_0, \psi$ simultaneously and extract Fourier 
coefficients indexed unambiguously by D(-1), D5 and NS5-brane charges. Instead,
one must decompose the action of  $N$ on functions
on $\cM_u$ into irreducible representations. By the Stone-von Neumann
theorem, any irreducible unitary representation of the Heisenberg algebra $[E_p,E_q]=E$ 
is either
\begin{itemize}
\item[i)]  one-dimensional,
with $E_p$ and $E_q$ acting as scalars and $E=0$, 
\item[iia)]  infinite-dimensional and  isomorphic to the action on 
the space of functions of two variables $x_0,y$ via
\be
\label{sw2}
E_p= \I x_0\ , \qquad E_q = -y\,  \pa_{x^0}\ ,\qquad E=\I y\ ,
\ee
\item [iib)] equivalently to iia) after Fourier transform, infinite-dimensional and isomorphic to the action on 
the space of functions of two variables $x_0,y$ via
\be
\label{sw1}
E_p= y\,  \pa_{x^0}\ ,\qquad E_q = \I x_0\ ,\qquad E=\I y\ .
\ee

\end{itemize}
In practice, this means 
that any function  $\Psi(t,\phi;\zeta,\tzeta,\sigma)$
 invariant under the Heisenberg group \eqref{heisper}
can be decomposed into its ``Abelian'' and ``non-Abelian'' parts\footnote{Such non-Abelian Fourier expansions have
been discussed in the mathematics literature for a variety of groups \cite{MR527787,MR1726680,MR2261342}. They also occur in condensed
matter physics in discussing Landau levels on the torus \cite{Onofri:2000hk}.
We are indebted to A. Neitzke  for numerous discussions on this subject.}, this means 
that any function  $\Psi(t,\phi;\zeta,\tzeta,\sigma)$
 invariant under the Heisenberg group \eqref{heisper}
can be decomposed into its ``Abelian'' and ``non-Abelian'' parts:
\be
\label{nonab}
\begin{split}
\Psi(t,\phi;\zeta,\tzeta,\sigma) &= \sum_{(p,q)\in\IZ^2}
\Psi_{p,q}(t,\phi)\, e^{2\pi \I (q \zeta - p \tzeta)}\\
&+ \sum_{k\in\IZ\backslash\{0\}} 
\sum_{\ell\in \IZ/(|k|\IZ)}\,
\Big[ 
\sum_{n\in\IZ+\frac{\ell}{|k|}} \, \Psi_{k,\ell}\left( t,\phi ; \tzeta - n \right)
\, e^{-2\pi \I k n \zeta-\pi \I k (\sigma- \zeta \tzeta)}  \Big] \ .
\end{split}
\ee
The first line is the contribution from one-dimensional type i) representations, and corresponds to the Fourier expansion of the constant term 
\be
\Psi(t, \phi, \zeta, \tzeta)=\int_0^2 \Psi(t, \phi, \zeta, \tzeta, \sigma) \de\sigma 
\ee
with respect to the ``Abelianized Heisenberg group" $\tilde{N}=N/Z$. 
Here $Z$ denotes the center of $N$ which coincides with the commutator subgroup $[N,N]$:
\be
Z=[N, N]=\left\{ \left(\begin{array}{ccc}
\ 1 & \ & \ *\  \\ 
 & 1 & \  \\
  & & 1 \\
  \end{array}\right) \right\}.
  \ee

The second line in (\ref{nonab}) then corresponds to infinite dimensional representations of type iia) with $y=-2\pi k$
and $x_0=-2\pi k n$. Note that the invariance of \eqref{nonab} under shifts $\zeta\mapsto \zeta+1, \sigma\mapsto \sigma+\tzeta$ is immediate since $\sigma-\zeta\tzeta$ is invariant and $n$ is integer; 
under shifts  $\tzeta\mapsto \tzeta+1, \sigma\mapsto \sigma-\zeta$ , the summation 
variable $n$ must be shifted, but the variation
of $\pi \I k (\sigma-\zeta\tzeta)$ and $2\pi \I k n$ in the exponential compensate each other,
so \eqref{nonab} is again invariant.

Equivalently, the same function may be decomposed into representations
of type i) and iia), as
\be
\label{nonab2}
\begin{split}
\Psi(t,\phi;\zeta,\tzeta,\sigma) &= \sum_{(p,q)\in\IZ^2}
\Psi_{p,q}(t,\phi)\, e^{2\pi \I (q \zeta - p \tzeta)}\\
&+ \sum_{k\in\IZ\backslash\{0\}} 
\sum_{\ell'\in \IZ/(|k|\IZ)}\,
\Big[ 
\sum_{m\in\IZ+\frac{\ell'}{|k|}} \, \tilde\Psi_{k,\ell'} \left( t,\phi; \zeta - m \right)
\, e^{2\pi \I k m \tzeta-\pi \I k (\sigma+ \zeta \tzeta)}  \Big] \ .
\end{split}
\ee
The relation between the two sets of non-Abelian Fourier coefficients follows
by Poisson resummation over $n$, and is given by Fourier transform, 
\be
\tilde\Psi_{k,\ell'}(t,\phi; \zeta) = \sum_{\ell=0}^{|k|-1} e^{-2\pi \I \frac{\ell \ell'}{k}}\, 
 \int_{-\infty}^\infty  \Psi_{k,\ell} (t,\phi; \tzeta) \, e^{2\pi i k \zeta \tzeta} \, \de\tzeta\ ,
 \label{changepol}
\ee
while the Abelian Fourier coefficients in \eqref{nonab} and \eqref{nonab2} are  
of course identical. Thus, the non-Abelian Fourier coefficients $\Psi_{k,l}$
and $\tilde\Psi_{k,l'}$ exhibit a wave function property, i.e. should really be 
thought of as a single state,
which can be expressed in different 
polarizations.\footnote{This property suggests that $\tilde\Psi_{k,\ell}$ may be closely
related to the topological string amplitude, or rather to its one-parameter generalization 
advocated in \cite{Gunaydin:2006bz}.}

Before we proceed to the more relevant case of the full principal series, as an example we give here the non-Abelian Fourier coefficients of the minimal Eisenstein 
series \eqref{eismin}, as computed in detail in Appendix \ref{App:MinimalFourier}:
\bea
\Psi_{0,q} &=&
\frac{2\, \pi^s}{\Gamma(s)} \, (e^{\phi}/t)^s \,
\mu_{2s-1}(q)
\, \cK_{s-\frac12}\left( 2\pi \,e^{\phi/2} \,t^{-3/2}\, |q| \right)\ , \nn\\
\Psi_{p,0} &=& \frac{2\,  \pi^s }{\Gamma(s)}\, 
(t\, e^\phi)^{\frac32-s} \, \mu_{2-2s}(p)
\, \cK_{1-s}\left( 2\pi \,e^{\phi/2} \,t^{3/2}\, | p |  \right)\ ,\nn\\
\Psi_{k,\ell}  &=&  \frac{2\, \pi^s  }{ \Gamma(s)}\,(e^{\phi}/ t)^s\,
\mu_{2s-1}(k,\ell)
\, \cK_{s-\frac12}\left( 2\pi \,  |k|  \,e^{\phi/2} \,\sqrt{  e^\phi  +  t^{-3}\, \tzeta ^2 }\right) \ ,\label{psimin}\\
\tilde\Psi_{k,\ell'}  &=& 
\frac{2\,\pi^s}{\Gamma(s)}\, (t\, e^{\phi})^{\frac32-s}\,
\mu_{2-2s}(k,\ell')
\,\cK_{1-s}\left( 2\pi \, |k| \,e^{\phi/2} \,\sqrt{ e^\phi  + \zeta ^2 t^3 } \right) \ ,\nn\\
\Psi_{0,0} &=&2 \zeta(2s) \, (e^{\phi}/t)^s\,  +
\frac{2\, \Gamma(s-\frac12)\,  \zeta(2s-1) }{\pi^{-\frac12}\Gamma(s) }e^{\frac{\phi}{2}} \,t^{2s-\frac32} 
+ \frac{2\, \Gamma(s-1)  \zeta(2s-2) }{\pi^{-1}\Gamma(s) }(t\,e^{\phi})^{\frac32-s}\ ,
\nn
\label{MinimalFourierCoefficients}
\eea
where the ``instanton measure'' for $n$ charges is generally defined by a sum over the common divisors of all charges
\be
\mu_s(N_1,\dots, N_n)\equiv \sum_{ m|N_1,\dots, N_n} |m|^{s}. 
\ee
These Fourier coefficients are considerably simpler than those arising from the principal Eisenstein
series $E(g;s_1,s_2)$ discussed in the next subsection, but they illustrate their general structure.  
In fact, they arise as  the limit $(s_1,s_2)\to (s,0)$ of a subset of the coefficients of $E(g;s_1,s_2)$.
Were this limit to describe some physical coupling, $\Psi_{0,p}$ and $\Psi_{q,0}$ would
correspond to  D5- and D$(-1)$ -instanton effects, with instanton actions displayed in (\ref{InstantonActionD5}) and (\ref{InstantonActionD(-1)}) below, while $\Psi_{k,\ell'}$ would
correspond to $(p,k)$ 5-branes with $p=km\in\IZ$ and vanishing D(-1) charge $Q=0$, 
as in \eqref{InstantonActionDD}. However, the principal Eisenstein
series $E(g;s_1,s_2)$ with $s_2\neq 0$ displays additional contributions
with $pq\neq 0$ and $Q\neq 0$, and considerably more involved instanton measure (\ref{NonAbelianMeasure}). A representation-theoretic point of view on the non-Abelian 
Fourier
expansions \eqref{nonab} and \eqref{nonab2} is provided in Appendix D.

\subsection{Generalized Eisenstein series and NS5-branes \label{secgeni}}

Let us now turn to the non-Abelian Fourier expansion of the full Eisenstein series 
$E(g;s_1,s_2)$ in \eqref{eisgen}. As it turns out, this  
was computed thirty years ago by Vinogradov and Takhtajan \cite{MR527787}
and by Bump \cite{MR765698}. 
In this section, we summarize their results adapted to our conventions, and we identify the 
instanton configurations responsible for each contribution. We work in terms of 
the variables $\{\nu, \tau_2, \tau_1, c_0, \psi\}$, which can be converted into 
the variables $\{t, \phi, \zeta, \tzeta, \sigma\}$ used in the previous section using  \eqref{chvar}. 
The main reason for this choice is that the variables $\tau=\tau_1+\I \tau_2$ 
and $(c_0,\psi)$ have simple transformation properties \eqref{sl2tau} under the S-duality group $SL(2,\mbb{Z})\subset SL(3,\mbb{Z})$, which is manifest in 
the Fourier expansion \cite{MR527787}.

\subsubsection{Constant terms}
As already discussed in Section \ref{Sec:ConstantTerms},
the term $\Psi_{0,0}$ in the Fourier expansion, depending only on 
the ``dilatonic" parameters $\nu,\tau_2$, corresponds to the 
``constant terms"  with respect to the Borel subgroup (\ref{Sec:ConstantTermMinimalParabolic}).
These terms are discussed in Appendix A, and agree with the analysis of  \cite{MR527787}:
\be
 \begin{split}
 \Psi_{0,0}(\nu, \tau_2; s_1, s_2) &= \int_{0}^{1} d\tau_1 \int_{0}^{1} d c_0 \int_{0}^{1} d\psi \ E(\nu, \tau_2, \tau_2, c_0, \psi; s_1, s_2)\\
 & = \nu^{-\frac{2s_1+s_2}{3}} \tau_2^{s_2}
 + c(s_1)  \,\nu^{-\frac12+\frac{s_1-s_2}{3}} \tau_2^{s_3}
 + c(s_2)  \,\nu^{-\frac{2s_1+s_2}{3}} \tau_2^{1-s_2}\\
&+ c(s_1)\,c(s_3)\,  \nu^{-\frac12+\frac{s_1-s_2}{3}} \tau_2^{1-s_3}
+c(s_2)\,c(s_3)\,  \nu^{\frac{2s_1+s_2}{3}-1} \tau_2^{s_1}\\
&+c(s_1)\,c(s_2)\,c(s_3)\,  \nu^{\frac{2s_1+s_2}{3}-1} \tau_2^{1-s_1}\ ,
\end{split}
 \ee
 where
 \be
 c(s)=\frac{\xi(2s-1)}{\xi(2s)}\ ,\qquad \xi(s)=\pi^{-s/2}\,\Gamma(s/2)\, \zeta(s)\ .
 \ee
 and $s_3=s_1+s_2-\frac12$.
 These terms were already discussed in \eqref{ep} for $(s_1,s_2)=(3/2,-3/2)$.
 We shall not discuss them any further here, except to note that they are consistent
 with the functional equations \eqref{weyls} obeyed by 
 the completed series \eqref{compEp}.

\subsubsection{Abelian Fourier coefficients}
Let us now proceed to analyze the Abelian Fourier coefficients $\Psi_{p,q}$ 
with $(p,q)\neq (0,0)$, starting with the simplest cases $\Psi_{0,q}$ and $\Psi_{p,0}$.
As shown in \cite{MR527787},  
\be
\begin{split}
\Psi_{0,q}(\nu, \tau_2)&=\phantom{+} \f{2(2\pi)^{1/2-s_1} c(s_2)c(s_3)}{\xi(2s_1)}\nu^{\f{s_1+2s_2}{3}-1} \tau_2^{1-s_1} \,\mu_{1-2s_1}(q) \, \, \mc{K}_{1/2-s_1}\big(2\pi |q| \tau_2\big)\\
&\quad+ \f{2(2\pi)^{1/2-s_2}}{\xi(2s_2)} \nu^{-\f{2s_1+s_2}{3}} \tau_2^{1-s_2}\,
\mu_{1-2s_2}(q) \,  \, \mc{K}_{1/2-s_2}\big(2\pi |q| \tau_2\big) \\
&\quad+ \f{2(2\pi)^{1/2-s_3}c(s_1)}{\xi(2s_3)} \nu^{\f{s_1-s_2-1}{2}} \tau_2^{1-s_3}\,
\mu_{1-2s_3}(q) \,\, \mc{K}_{1/2-s_3}\big(2\pi |q| \tau_2\big)\ ,
\end{split}
\ee
and 
\be
\begin{split}
\Psi_{p, 0}(\nu, \tau_2)&=\phantom{+} \f{2(2\pi)^{1/2-s_1}}{\xi(2s_1)} \nu^{-\f{2s_1+s_2}{3}}
\tau_2^{s_2}\mu_{1-2s_1}(p) \,  \, 
\mc{K}_{1/2-s_1}\big(2\pi |p|/\sqrt{ \nu \tau_2} \big)\\
&\quad+ \f{2(2\pi)^{1/2-s_2}c(s_1)c(s_3)}{\xi(2s_2)} 
\nu^{\f{s_1-s_2}{3}-\f{1}{2}}\tau_2^{\f{3}{2}-s_1-s_2}\, 
\mu_{1-2s_2}(p) \,  \, \mc{K}_{1/2-s_2}\big(2\pi |p| /\sqrt{ \nu \tau_2}\big)\\
&\quad+ \f{2(2\pi)^{1/2-s_3}c(s_2)}{\xi(2s_3)} \nu^{-\f{s_2+2s_1}{3}}
\tau_2^{1-s_2} \, \mu_{1-2s_3}(p) \, \, \mc{K}_{1/2-s_3}\big(2\pi |p| /\sqrt{ \nu \tau_2}\big)\ .
\end{split}
\ee
By comparing with (\ref{MinimalFourierCoefficients}) it is apparent that these have a very similar structure to the minimal Eisenstein series. Using the asymptotic expansion of the Bessel function \eqref{asK}, we deduce that in the weak coupling limit $\tau_2\to\infty$, the coefficients $\Psi_{0,q}$ contribute to the expansion (\ref{nonab2}) by exponentially suppressed contributions of order $e^{-2\pi S_{0,q}}$, with 
\be
S_{0,q}(\tau)=|q| \, \tau_2+\I \, q\, \tau_1.
\label{InstantonActionD(-1)}
\ee
This is precisely the instanton action for D$(-1)$ instantons \cite{Gibbons:1995vg,Green:1997tv}. Similarly, the coefficients $\Psi_{p,0}$ encode D$5$-brane instantons, with classical action
\be
S_{p,0}(\nu, \tau_2, c_0)=|p| (\nu \tau_2)^{-1/2}-\I \, p \, c_0 = |p|\, \tau_2 V \,  -\I \, p\, c_0 \ .
\label{InstantonActionD5}
\ee
From \eqref{asK} of the Bessel function, we may also extract the ``instanton measure" $\mu(p,q)$, defined by 
\be
\Psi_{p, q}(\nu,\tau_2) \sim \tau_2^\alpha\,  \nu^\beta\, \big(\Re\ S_{p,q}\big)^\gamma \, \mu(p,q) \, e^{-2\pi (\Re\ S_{p,q})}
\label{defmes}
\ee
in the weak coupling limit $\tau_2\to \infty$, with suitable choices of $\alpha,\beta,\gamma$
to absorb the moduli dependence. The prefactors in front of $e^{-2\pi S_{p,q}}$, including the instanton
measure, should arise from the external vertices and the fluctuation determinant in the instanton background \cite{Green:1997tv}. For the particular case of D(-1)-D5 instantons with $pq=0$, we find
\be
\mu(0,q) \equiv \f{c(s_2)c(s_3)}{\xi(2s_1)} \mu_{1-2s_1}(q) +\f{1}{\xi(2s_2)} \mu_{1-2s_2}(q)+\f{c(s_1)}{\xi(2s_3)}  \mu_{1-2s_3}(q).
\ee
and 
\be
\mu(p,0) \equiv \f{1}{\xi(2s_1)}  \mu_{1-2s_1}(p)+\f{c(s_1)c(s_3)}{\xi(2s_2)}\mu_{1-2s_2}(p)+\f{c(s_2)}{\xi(2s_3)} \mu_{1-2s_3}(p).
\ee

We now proceed to analyze the coefficients $\Psi_{p,q}$ for $pq\neq0$, which we recall were absent for the minimal Eisenstein series. These coefficients may be written as
\be
\label{genab}
\begin{split}
\Psi_{p,q}(\nu, \tau_2)&=\f{4\nu^{\f{s_2-s_1}{6}-\f{1}{2}} \tau_2^{\f{s_2-s_1}{2}+\f{1}{2}}}{\xi(2s_1)\xi(2s_2)\xi(2s_3)} 
 \sum_{d_1|p} \sum_{d_2|\frac{p}{d_1}} 
d_1^{1-2s_3} d_2^{1-2s_2} \\
& \qquad \times \sigma_{1-2s_1,1-2s_3}\Big(\f{p}{d_1d_2}, |q|\Big) 
\, (pq)^{s_3-\frac12} \,  \mc{I}_{s_1,s_2}\left( R_{p,q}, \x_{p,q} \right)\ ,
\end{split}
\ee
where
\be
R_{p,q}\equiv \left(\f{p^2 |q|}{\nu}\right)^{1/3}, \qquad \x_{p,q}\equiv \tau_2^{-1} \left(\frac{p^2}{\nu q^2}\right)^{1/3} = t^2  \left(\frac{|p|}{|q|}\right)^{2/3}\ .
\label{pqvariables}
\ee
Here, we have also defined the ``double divisor sum''
\be 
\sigma_{\alpha,\beta}(n,m) \equiv \sum_{m=d_1 d_2 d_3,\atop
d_1,d_2,d_3>0, \gcd(d_3,n)=1} d_2^{\alpha} d_3^{\beta}\ ,
\ee
and the integral
\be
\label{integral}
 \mc{I}_{s_1,s_2}(R,\x)
\equiv  \int_0^{\infty} K_{s_3-1/2}\big(2\pi R\,  \x^{-1}  \sqrt{1+x}\big) \,K_{s_3-1/2}\Big(2\pi R \, \x^{1/2} \sqrt{1+1/x}\Big) \, x^{\f{s_2-s_1}{2}} \f{dx}{x}\ .
 \ee
  At weak 4D coupling $\nu\to 0$ keeping $t$ fixed, 
 one may use the saddle point approximation of the integral
 (for details, see Appendix \ref{SaddlePoint})
\be
\label{sadasI}
\mc{I}_{s_1,s_2}(R,\x) \sim 
\f{\x^{\f{s_2-s_1+1}{2}}  }{\sqrt6\, R^{3/2} (1+\x)^{1/4}}  \exp\left[- \frac{ 2\pi R(1+\x)^{3/2}}{\x}\right]
\left( 1 + \frac{I_1}{R}+ \mathcal{O}(1/R^2) \right)\ .
\ee
Plugging in the values of $R_{p,q}$ and $\x_{p,q}$ given in \eqref{pqvariables}, we find that such 
terms give exponentially suppressed contributions of order $e^{-2\pi S_{p,q}}$, where
\be
\label{InstantonActionDD}
S_{p,q}(\nu, \tau, c_0)=\Big[(\tau_2V |p|)^{2/3}+( \tau_2 |q|)^{2/3}\Big]^{3/2}+\I( q\tau_1-p c_0)\ .
\ee
We note that the real part of this action is proportional to the mass formula for bound states of D0-D6-branes found in \cite{Dhar:1998ip,Rasheed:1995zv,Larsen:1999pp}. Moreover, in the limit $q=0$ or $p=0$, 
\eqref{InstantonActionDD} reduces to (\ref{InstantonActionD(-1)}) and (\ref{InstantonActionD5}). 
We conclude that general
Abelian terms with $pq\neq 0$ correspond to bound states of D(-1) and D5-brane instantons. 
Their  summation measure, defined as in \eqref{defmes}, is given by
\be
\mu(p,q)\equiv 
\f{4}{\sqrt{6}}\f{\, |p|^{\f{2s_1+4s_2}{3}-\f{5}{6}}|q|^{\f{4s_1+2s_2}{3}-\f{5}{6}} }{\xi(2s_1)\xi(2s_2)\xi(2s_3)} 
\sum_{d_1|p} \sum_{d_2|\frac{p}{d_1}}  
d_1^{1-2s_3} d_2^{1-2s_2}
\sigma_{1-2s_1,1-2s_3}\Big(\f{|p|}{d_1 d_2}, |q|\Big)\ .
\label{InstantonMeasureDD}
\ee
According to the general relation between instanton measure and BPS black hole degeneracies
proposed in \cite{Alexandrov:2008gh}, the measure \eqref{InstantonMeasureDD} should be related to the D0-D6 bound state degeneracies, which are known to be encoded in  the Mac Mahon function \cite{Denef:2007vg}. This would provide a crucial check of our proposal, which we leave for future work. 

On the type IIA side, the Abelian terms $(p,q)$  correspond
to D2-branes wrapped on the 3-cycle $p\,\gamma_0-q\, \gamma^0$,
where $\gamma_0$ and $\gamma^0$ are the 3-cycles singled out by the
large complex structure limit.

\subsubsection{Non-Abelian Fourier coefficients}
Following \cite{MR527787}, the general non-Abelian Fourier coefficients are given by 
\be
 \label{nonAbeliancoefficient}
\begin{split}
\tilde\Psi_{k,\ell'}(\nu,\tau) &=
\f{4\nu^{\f{s_2-s_1}{6}-\f{1}{2}} }{\xi(2s_1)\xi(2s_2)\xi(2s_3)} 
\sum_{q\in\IZ} \sum_{d_1|d} \sum_{d_2|\frac{d}{d_1}} 
d_1^{1-2s_3} d_2^{1-2s_2} 
\sigma_{1-2s_1,1-2s_3}\Big(\f{d}{d_1d_2}, |q|\Big) \,
\\
&\quad\times [\tau_2]_{-k,p}^{\f{s_2-s_1}{2}+\f{1}{2}} \, (d\,q)^{s_3-\frac12}\,
\mc{I}_{s_1,s_2}\left( R_{d,q},\x_{d,q} \right)\  e^{-2\pi \I q [\tau_1]_{-k, p}}  ,
\end{split}
\ee
where the notations are as follows:  $\mc{I}_{s_1,s_2}(R, \x)$ is defined in \eqref{integral},
$d\equiv\gcd(p,k)>0$,
\be
\label{newpq}
R_{d,q}\equiv \left(\f{d^2 |q|}{\nu}\right)^{1/3}, \qquad \x_{d,q}\equiv [\tau_2]_{-k,p} ^{-1} 
\left(\frac{d^2}{\nu q^2}\right)^{1/3} \ ,
\ee
and the variables
$[\tau_1]_{-k, p}$ and $[\tau_2]_{-k, p}$
denote the real and imaginary parts of the image of $\tau=\tau_1+\I \tau_2$ under an $SL(2,\IZ)$
transformation of the form \eqref{SL2Z},
\be
\delta=\left(\begin{array}{cc}
\al & \beta \\
-k^{\prime} & p^{\prime} \\
\end{array}\right)\ ,\qquad
\delta\cdot \tau=
\f{\al \tau+\beta}{-k^{\prime} \tau+p^{\prime}}\equiv [\tau_1]_{-k,p}
+\I[\tau_2]_{-k,p},
\label{deft12}
\ee
where $k^{\prime}=k/d$ and $p^{\prime}=p/d$  and 
$\alpha,\beta$ are two integers such that $\alpha p'+\beta k'=1$. Since $k\neq 0$, 
this is usefully rewritten as
\be
[\tau_2]_{-k, p}=\f{d^2 \tau_2}{|p-k\tau|^2}, \qquad [\tau_1]_{-k, p}=-\frac{d \alpha}{k}+
\f{d^2 (p-k\tau_1)}{k|p-k\tau|^2}\ ,
\label{vtransformed}
\ee
where we used $\beta=(1-p^{\prime}\al)/k^{\prime}$ to derive the second relation. 
Defining $Q\equiv d^2 q$, \eqref{newpq} may therefore be rewritten as
\be
\label{newpq2}
R_{d,q}= \tau_2 t^2 |Q|^{1/3}\ , \qquad \x_{d,q}=
\frac{t^2\, |p-k\tau|^2}{|Q|^{2/3}}\ .
\ee
In type IIA variables suitable for comparison with (\ref{nonab2}), this becomes
\be
\label{newpq3}
R_{d,q}= e^{\phi/2} t^{1/2} k n \ ,\qquad \x_{d,q}=
\frac{e^{\phi}+ t^3 (m-\zeta)^2}{n^2 t} \ ,
\ee
where we have use \eqref{chvar} and defined
\be
m = \frac{p}{k}   \in \mbb{Z}+ \f{\ell^{\prime}}{|k|}\ ,\qquad n = \sgn(Q) \, |Q|^{1/3} / k\ .
\ee
In particular, \eqref{nonAbeliancoefficient} depends only on the difference $\zeta-m$.
Setting $m=0$, we therefore have
\be
 \label{nonAbeliancoefficient2}
\begin{split}
\tilde\Psi_{k,\ell'}(t,\phi;\zeta) &=
\f{4 t^{\f{s_2-s_1+3}{2}} e^\phi d^{2(1-s_1)}}{\xi(2s_1)\xi(2s_2)\xi(2s_3)} 
 \sum_{n:\, (k n)^3\in d^2\IZ}
 \sum_{d_1|d} \sum_{d_2|\frac{d}{d_1}} 
d_1^{1-2s_3} d_2^{1-2s_2} 
\sigma_{1-2s_1,1-2s_3}\Big(\f{d}{d_1d_2}, |q|\Big) \,
\\
& \left(e^{\phi}+ t^3 \zeta^2\right)^{\f{s_1-s_2-1}{2}} 
\, |kn|^{s_3-\frac12}\, \,
\mc{I}_{s_1,s_2}\left( e^{\phi/2} t^{1/2} k n,\frac{e^{\phi}+ t^3 \zeta^2}{n^2 t}\right)\  
e^{-  \I k \frac{t^3 n^3 \zeta}{e^\phi+t^3 \zeta^2} + 2\pi \I \frac{k^2 n^3 \alpha}{d}} .
\end{split}
\ee
where now $d=\gcd(\ell',k)$.

As before, using the saddle point approximation of the integral $\mc{I}(R, \x)$ (see appendix \ref{SaddlePoint}), we find that $\tilde{\Psi}_{k,\ell^{\prime}}$ contributes to (\ref{nonab2}) with terms which are exponentially suppressed by $e^{-2\pi S_{Q, p, k}}$, where
\be
\label{genS}
S_{Q, p, k}
= \f{\Big[(V\tau_2)^{2/3}|p-k\tau|^2+\tau_2^{2/3} Q^{2/3}\Big]^{3/2}}{|p-k\tau|^2}
+\I \, Q  \f{p-k\tau_1}{k|p-k\tau|^2} -\I (p\,c_0+k\,\psi) \ ,
\ee
or, in type IIA variables,
\be
\label{Skl1}
S_{m,n,k}= |k| \,\frac{e^{\phi/2} [e^\phi+t^3 (\zeta-m)^2 +t n^2]^{3/2}}{e^\phi+ t^3 (\zeta-m)^2 } 
-  \I k \frac{t^3 n^3 (\zeta-m)}{e^\phi+t^3 (\zeta-m)^2} 
+\frac{\I}{2} k (\sigma+\zeta\tzeta) - \I k m \tzeta \ .
\ee 
In these expressions, the last two terms originate from the phase factors 
appearing in the Fourier expansion (\ref{nonab2}).
From the saddle point approximation \eqref{sadasI}, we may also extract the  summation 
 measure defined as in \eqref{defmes},
\be
\mu(Q,p,k)\equiv 
\f{4 e^{\f{2\pi \I Q  \al }{d k}}}{\sqrt{6}}\f{\, |d|^{\f{5}{6}-2s_1}|Q|^{\f{4s_1+2s_2}{3}-\f{5}{6}} }{\xi(2s_1)\xi(2s_2)\xi(2s_3)} 
\sum_{d_1|d} \sum_{d_2|\frac{d}{d_1}}  
d_1^{1-2s_3} d_2^{1-2s_2}
\sigma_{1-2s_1,1-2s_3}\Big(\f{|d|}{d_1 d_2}, |Q/d^2|\Big)\ .
\label{NonAbelianMeasure}
\ee
Contrary to appearances, the limit $k\to 0$ is smooth: the apparent singularity
cancels between the action and the summation measure, as the two
terms in \eqref{vtransformed} combine into
\be
 [\tau_1]_{-k, p}=
 \frac{d[(\alpha p - \beta k)\tau_1 + \beta p - \alpha k  |\tau|^2]}{|p-k\tau|^2} 
\stackrel{k\to 0}{\longrightarrow}
\frac{d}{|p|} (\alpha\tau_1-\beta)= 
\tau_1 \ ,
\label{vtransformedcomb}
\ee
where in the last step we used the fact that $\alpha=1,\beta=0,d=|p|$ when $k=0$.
In this limit we therefore recover the action \eqref{InstantonActionDD} of D(-1)-D5 bound states. Moreover, the term with $k\neq 0$ can be recovered from $(Q,p,k)=(d^2 q,d,0)$ by an $SL(2,\IZ)$ action.
It would be interesting to recover  \eqref{Skl1} from the type IIA five-brane
action \cite{Witten:1996hc}, however by duality it is clear that \eqref{Skl1} must describe the
action of $k$ NS5-branes bound to $km$ D2-branes wrapped
on $\gamma_0$ and $(kn)^3$ D2-branes wrapped on $\gamma^0$.

Therefore, we conclude that \eqref{nonAbeliancoefficient} describes the contribution of general bound states 
of  $Q$ D$(-1)$-instantons and $(p,k)$ 5-branes (equivalently, on the type IIA side,   
$NS5-D2$ bound states).
Thus, despite the fact that the non-Abelian
nature of the Heisenberg group prevents us from defining D$(-1)$ and D$5$ brane charges
unambiguously when the NS5-brane charge $k$ is non-zero, we still find that the general
term involves a contribution of D(-1), D5 and NS5-brane instantons, with independent charges
$Q,p,k$.

\subsection{The minimal theta series as a NS5-brane partition function}

The general non-Abelian contribution \eqref{genS} will undoubtedly remind the cognoscente
of the minimal representation of $G_{2(2)}$ constructed 
in Sec. 3.5. of \cite{Gunaydin:2007qq}. Indeed, under the standard embedding $SL(3,\IR)
\subset G_{2(2)}$, the minimal representation of $G_{2(2)}$ belongs 
to the non-spherical supplementary series of $SL(3,\IR)$ with infinitesimal parameters $(s_1,s_2)=(2/3,2/3)$ \cite{MR1253210}; although this representation 
admits no spherical vector, its lowest K-type is indistinguishable from the 
spherical vector of the principal representation in the strict classical limit. 
Moreover, the minimal representation of $G_{2(2)}$ is a special instance
of the minimal representation of the quasi-conformal group
${\rm QConf}(J)$ attached to any cubic Jordan algebra $J$, in the case where $J=\IR$,
$\kappa_{111}=6$. These
minimal representations were constructed in \cite{Kazhdan:2001nx}
for simply-laced Lie groups in the split real form (albeit not using the
language of Jordan algebras), and more recently for all simple 
Lie groups in any non-compact real form, in particular the quaternionic real form, 
in \cite{Gunaydin:2004md,Gunaydin:2005zz,Gunaydin:2006vz}. The latter can all
be reached by analytic continuation from 
the minimal representation for the split real form constructed in \cite{Kazhdan:2001nx}.
Their lowest K-type is not known in general yet, but from the $G_{2(2)}$ example it is 
clear that it will be identical in the strict classical limit to the spherical vector for the split real form,
found in \cite{Kazhdan:2001nx} and displayed in \eqref{sphming} below.

Having recalled this representation-theoretic background, we now 
assume that the HM moduli space $\cM_K(Y)$ is given at tree
level  by the QK symmetric space \eqref{MJ}, and that a larger 
discrete symmetry $G(\IZ)={\rm QConf}(J,\IZ)\supset SL(3,\IZ)$
remains unbroken by quantum corrections, and argue that the
minimal theta series of $G(\IZ)$, i.e. the automorphic form attached to the minimal representation\footnote{Early suggestions
that the minimal representation is relevant for black hole counting were made in
\cite{Gunaydin:2000xr,Gunaydin:2001bt,Pioline:2005vi,Gunaydin:2006bz}.}
via \eqref{overl}, predicts exponential corrections of the form \eqref{genS}, where the D-instanton 
charge $Q$ is now a composite of D3-brane charges $N^a$ labelling
the various 4-cycles in $Y$.

To make this more precise, we use the expression \eqref{sphming} for the approximate spherical vector $f_K$, and incorporate the moduli dependence by acting with $\rho(e^{-1})$, where $e\in SO(4) \backslash G_{2(2)}$ is the Iwasawa coset representative in  \eqref{iwag2}
and $\rho$ is the minimal representation obtained in Section 3.5.1 of \cite{Gunaydin:2007qq}  
(or rather, its Fourier transform over $(x_0,x_1)\mapsto (x^0,x^1)$). By further relabeling $y=2\pi k,x^0=2\pi p, x^1 = 2\pi N^1$, we then find that the minimal theta series of $G_{2(2)}$ \eqref{overl} predicts exponentially suppressed contributions of order $e^{-2\pi S}$ 
where 
\be
\begin{split}
S
&= \tau_2 V\, 
|p-k\tau|\, \left( 1 + \frac{ (\Nt^1/t)^2}{|p-k\tau|^2}
\right)^{3/2}
 -\I (p c_0 + k \psi + N^1 c_1) 
+\I \,\frac{(p-k \tau_1) }{k |p-k\tau|^2} (\Nt^1)^3\\& \qquad\qquad 
\hspace*{-2mm}
+\frac{3\I}{k}  \, b^1 (\Nt^1)^2
 + \frac{\I}{k} \, b^1 \, ( p b^1-k c^1)\, (3 \Nt^1+p b^1 - k c^1)\ ,
\label{SmingG2}
\end{split}
\ee
where $\Nt^1 = N^1 + k c^1- p b^1$.
More generally, for an arbitrary quasi-conformal group $G={\rm QConf}(J)$, 
the same procedure based on \eqref{sphming}  and the results in 
\cite{Gunaydin:2004md,Gunaydin:2005zz} leads to\footnote{We are grateful to S. Alexandrov
for pointing out an error in the previous version of this formula.}
%\be
%\begin{split}
%S
%&= \tau_2 V\, 
%|p-k\tau|\, \left( 
%1 + 3 \frac{ (\Nt^a/t^a)^2}{|p-k\tau|^2}
%+ \frac1{12} \frac{  (\kappa_{abc} t^a \Nt^b \Nt^c)^2}{V^2|p-k\tau|^4}
%+ \frac1{2^2\cdot 3^2} \frac{ (\kappa_{abc} \Nt^a \Nt^b \Nt^c)^2}{V^2|p-k\tau|^6}
%\right)^{1/2}\\
%& \qquad\qquad 
%-\I (p c_0 + k \psi + N^a c_a) 
%+\frac{\I}{6}  \,\frac{(p-k \tau_1) }{k |p-k\tau|^2} \kappa_{abc}\Nt^a \Nt^b \Nt^c \\
%& \qquad\qquad+\frac{\I}{2k}  \kappa_{abc} \, b^a \Nt^b \Nt^c
% - \frac{\I}{6k}  \kappa_{abc} \, b^a \, ( p b^b-k c^b)\, (3 \Nt^c+p b^c - k c^c)
%\ ,
%\label{Sming}
%\end{split}
%\ee
\be
\begin{split}
S=&\tau_2 V\,
|p-k\tau|\, \left[
1 +  \frac{(t, t, \Nt)^2 -4 V (t,\Nt ,\Nt)}{4V^2|p-k\tau|^2}
%\right.
%\\
%& \qquad\qquad\qquad \left.
+ \frac{  (t, \Nt ,\Nt)^2 - \frac23 (\Nt,\Nt,\Nt)(t t\Nt)}{4V^2|p-k\tau|^4}\right.\\
& \qquad \left.
+  \frac{ (\Nt,\Nt,\Nt)^2}{2^2 3^2\,V^2|p-k\tau|^6}
\right]^{1/2}
-\I (p c_0 + k \psi + N^a c_a)
+\I  \,\frac{(p-k \tau_1) }{6k |p-k\tau|^2} (\Nt,\Nt,\Nt )\\
&+\frac{2\pi\I}{2k}  (b,\Nt, \Nt)
 + \frac{\I}{6k}  (b , p b-k c,3 \Nt+p b - k c)\, ,
\label{Sming}
\end{split}
\ee
where $\Nt^a = N^a + k c^a- p b^a$ and $(x,y,z)\equiv\kappa_{abc} x^a y^b z^c$. 
Here and in \eqref{SmingG2},
the type IIB variables $(c^a,c_a,c_0,\psi)$ are related to $(\zeta^a,\tzeta_a,\tzeta_0,\psi)$
by the tree-level mirror map, Eq. (3.20) in \cite{Alexandrov:2008gh}.
Of course  \eqref{Sming} reduces to \eqref{SmingG2}
upon setting $\kappa_{111}=6$.

Eq. \eqref{Sming} is recognized as the action of a $(p,k)$5-brane bound to 
D3-branes wrapping $N^a \gamma_a \in H_4(Y,\IZ)$, with induced 
D1-brane charge $\tilde N_a$ and D(-1)-instanton charge $\tilde Q$ given by
\be
\label{indcharge}
\tilde N_a=\frac1{6|p-k\tau|} \kappa_{abc} \Nt^b\Nt^c\ ,\qquad
\tilde Q= \frac1{6|p-k\tau|^2} \kappa_{abc} \Nt^a \Nt^b \Nt^c\ .
\ee
In particular, by the same token as in \eqref{vtransformed}, 
the last term in \eqref{Sming} is just the axionic coupling of $\tilde Q$ D-instantons
for vanishing NS5-brane charge, after rotating to the $(p,k)$5-brane duality frame.
It would be interesting to compare \eqref{Sming}
with other studies of NS5-instanton corrections to the HM moduli space
based on supergravity (see e.g. \cite{deVroome:2006xu,Chiodaroli:2009cz}
and references therein).

In the limit $k\to 0$, \eqref{Sming} reduces to the usual action of a D5-D3-D1-D(-1)
bound state, 
\be
S= \tau_2 \, 
\left( p^2 V^2+ 3 (\Nt^a/t^a)^2 V^2+ 3 (\Nt_a t^a)^2 + \tilde Q^2\right)^{1/2}
-\I (p c_0 + N^a c_a) 
\ ,
\label{Smingab}
\ee
with charges $(p,\Nt^a,\Nt_a,\tilde Q)$ given by 
\be
\Nt^a= N^a - p \, b^a\ ,\qquad
\tilde N_a=\frac1{6|p|} \kappa_{abc} \Nt^b \Nt^c \ ,\qquad
\tilde Q= \frac1{6|p|^2} \kappa_{abc} \Nt^a \Nt^b \Nt^c\ .
\ee
However, this is not the most general  
D5-D3-D1-D(-1) instanton correction, since the D1 and D(-1) charges are determined in terms of the
D5 and D3 charges $(p,N^a)$. This reflects the fact that 
the Abelian Fourier expansion of the minimal theta series has support
on ``very small'' charges satisfying $I_4=\pa I_4=\pa^2 I_4=0$, where $I_4$
is the quartic invariant of the 4D duality group ${\rm Conf}(J)$ (these conditions 
generalize the condition
$pq=0$ for the minimal representation of $SL(3,\IR)$). Therefore, one
should probably look for automorphic forms in the 
quaternionic discrete series \cite{MR1421947,Gunaydin:2007qq,Gunaydin:2009dq},
where this restriction does not apply. It is 
also  conceivable that only ``very small" charges may contribute to the hypermultiplet metric,
but this does not seem to be required by supersymmetry.

Nevertheless, it is tempting to conjecture that, in cases where the discrete symmetry 
$G(\IZ)={\rm QConf}(J,\IZ)$ is unbroken quantum mechanically, the minimal 
theta series of $G$, or an automorphic form in the quaternionic discrete series of $G$, 
may encode the effects of bound states of NS5-brane and D-instantons
 on the hypermultiplet moduli space. If so, it should be possible to express them as 
a sum of $SL(3,\IZ)$ invariant contributions as 
in \eqref{phiinvsl3} and compute the corresponding invariants $n^{(0)}_{k_a}$. 
Clearly, more work remains to establish this claim. For example, 
the minimal representation is naturally understood as a submodule 
of $H^1\big(\cZ_\cM,\cO(-(h+3)/3)\big)$ \cite{MR1421947},
whereas deformations of the \qk space $\cM$ are usually controlled by
$H^1\big(\cZ_\cM,\cO(2)\big)$ \cite{lebrun1988rtq,Alexandrov:2008nk}. Nevertheless, 
this proposal meshes well with ideas expressed in 
\cite{Gunaydin:2006bz}, where the minimal representation was related to the
topological string amplitude, and in \cite{Dijkgraaf:2002ac,Kapustin:2004jm}, where the topological
amplitude was related to the NS5-brane and D5-brane partition functions. It would also be 
interesting to extend these considerations to Calabi-Yau threefolds which
do not have a symmetric moduli space at tree level.

\acknowledgments

We are indebted to S. Alexandrov, L. Bao, M. Gutperle,
N. Halmagyi, A. Kleinschmidt,  S. Miller, B. E. W. Nilsson, C. Petersson, F. Saueressig,
L. Takhtajan,
S. Vandoren and P. Vanhove for useful discussions or correspondence.
Special thanks are due to A. Neitzke for 
discussions on NS5-branes, non-Abelian Fourier expansions and representation theory, 
and to A. Waldron for collaboration on $SL(3,\IZ)$ Eisenstein series in the past.
This research  is supported in part by ANR(CNRS-USAR) contract no. 05-BLAN-0079-01. 
\vskip 5mm 

%\noindent Note added in v2: 
%BP is grateful to DP for his remarks on the first, single authored
%version of this work, which led us, via \cite{MR527787}, to the Fourier expansion 
%of the principal Eisenstein series of $SL(3,\IZ)$. We hope that this and other
%improvements have made our case even more compelling. 

%\noindent {\bf\large Note added in v2}

%\vskip 5mm 

%After the first version of this work was released, I became aware of \cite{MR527787}, where the 
%non-Abelian Fourier expansion of the principal Eisenstein series \eqref{eisgen} is
%computed. This expansion is reproduced and adapted to the notations of this paper in Appendix C,
%written in collaboration with Daniel Persson.

\appendix

\section{Constant terms for $SL(3)$ Eisenstein series}
\label{ConstantTerms}

In this Section, we summarize the constant terms of the  Eisenstein 
series \eqref{eisgen} and \eqref{eismin} along the parabolic subgroups \eqref{defpmin} and \eqref{defpmax},  based on the general results of 
Langlands  \cite{MR0579181,MR0419366}, or the explicit computations 
in  \cite{MR527787,MR765698}. Our notation follows \cite{Pioline:2004xq}.

In general, the constant term of a $\Gamma$-invariant function $E(g)$ on $G$ 
with respect to a parabolic subgroup $P\subset G$ is defined by the integral
\be
E_{P}(g) = \int_{N/ [\Gamma\cap N]} \de n\, E(g)\ , 
\ee
where $N$ is the unipotent radical of $P$. These coefficients provide the ``perturbative" 
part of the automorphic 
form ($E(g)$  is said to be cuspidal if its constant terms $E_{P}(g)$ vanish for
all parabolic subgroups).

For the principal  Eisenstein series defined in \eqref{eisgen}, 
the constant term along the minimal parabolic \eqref{defpmin} is a sum
of six terms,
\be
\label{eisct}
\begin{array}{ccccc}
E_{P_{\rm min}}(g;s_1,s_2)&=&
 t_1^{\lambda_1+1}t_2^{\lambda_2}t_3^{\lambda_3-1}
&+&
 t_1^{\lambda_2+1}t_2^{\lambda_1}t_3^{\lambda_3-1}\
 \frac{\xi(\lambda_{12})}{\xi(\lambda_{21})}
\\[3mm]&&
 t_1^{\lambda_1+1}t_2^{\lambda_3}t_3^{\lambda_2-1}\
 \frac{\xi(\lambda_{23})}{\xi(\lambda_{32})}
&+&
 t_1^{\lambda_3+1}t_2^{\lambda_1}t_3^{\lambda_2-1}\
 \frac{\xi(\lambda_{13})}{\xi(\lambda_{31})}
 \frac{\xi(\lambda_{23})}{\xi(\lambda_{32})}
\\[3mm]&&
 t_1^{\lambda_2+1}t_2^{\lambda_3}t_3^{\lambda_1-1}\
 \frac{\xi(\lambda_{12})}{\xi(\lambda_{21})}
 \frac{\xi(\lambda_{13})}{\xi(\lambda_{31})}
&+&\, 
 t_1^{\lambda_3+1}t_2^{\lambda_2}t_3^{\lambda_1-1}\
 \frac{\xi(\lambda_{12})}{\xi(\lambda_{21})}
 \frac{\xi(\lambda_{13})}{\xi(\lambda_{31})}
 \frac{\xi(\lambda_{23})}{\xi(\lambda_{32})}
\end{array}\ .
\ee
Here, $\lambda_1,\lambda_2,\lambda_3$ are defined by
\be
\label{sla}
2s_1= 1+\lambda_2-\lambda_3\, ,\qquad
2s_2= 1+\lambda_1-\lambda_2\, ,\qquad
\lambda_1+\lambda_2+\lambda_3=0\ ,
\ee
$\lambda_{ij}$ denotes the difference $\lambda_i-\lambda_j$, 
the real variables $t_1,t_2,t_3$ are related to the Abelian part of 
the Iwasawa decomposition \eqref{iwa1} via
\be
(t_1,t_2,t_3) = (\nu^{-1/3},  \nu^{1/6} \sqrt{\tau_2},  \nu^{1/6}/\sqrt{\tau_2})
=(e^{\phi/2}\sqrt{t},1/t,e^{-\phi/2}\sqrt{t})\ ,  
\ee
and 
\be
\xi(s)=\pi^{-s/2}\,\Gamma(s/2)\, \zeta(s)
\ee
is the ``completed" Riemann Zeta function, satisfying $\xi(s)=\xi(1-s)$.
The Weyl group acts by permutations on $(\lambda_1,\lambda_2,\lambda_3)$,
and leaves invariant $\xi(\lambda_{21}) \xi(\lambda_{31})\xi(\lambda_{32})
E_{P_{\rm min}}(g;s_1,s_2)$. This invariance extends to the ``completed" 
Eisenstein series 
\be
\label{compEp}
\cE(g;\lambda_i) = \xi(\lambda_{21}) \, \xi(\lambda_{31}) \, \xi(\lambda_{32})\, 
E(g;s_1,s_2)\ ,
\ee
which is invariant  under permutations of $(\lambda_1,\lambda_2,\lambda_3)$,
corresponding in the $(s_1,s_2,s_3\equiv s_1+s_2-\frac12)$ variables to
\be
\label{weyls}
(s_1,s_2) \sim (1-s_2,1-s_1) \sim (1-s_1,s_3) \sim (s_3,1-s_2)
\sim (1-s_3,s_1)\sim(s_2,1-s_3).
\ee
On the other hand, the constant terms around the maximal parabolic \eqref{defpmax}
are given by 
\bea
\label{minpmax}
E_{P_{\rm max}}(g;s_1,s_2)&=&t_1^{\lambda_{12}-\frac12\lambda_{32}+\frac32}\ 
\cE_{\lambda_{32}}(\tau)
+\frac{\xi(\lambda_{12})}{\xi(\lambda_{21})}
t_1^{\lambda_{21}-\frac12\lambda_{31}+\frac32}
\cE_{\lambda_{31}}(\tau)\nn\\
&+&
\frac{\xi(\lambda_{23})\xi(\lambda_{13})}{\xi(\lambda_{32})\xi(\lambda_{31})}
t_1^{\lambda_{31}-\frac12\lambda_{21}+\frac32}
\cE_{\lambda_{21}}(\tau)\, ,
\label{p2}
\eea
where $\tau=a_3+i(t_2/t_3)$. 
Here $\cE_s(\tau)$ is related to the 
$SL(2,\IZ)$ Eisenstein series \eqref{sl2eis} via
$\cE_{1-2s}(\tau)= E_s(\tau) / (2\zeta(2s))$,
%\be
%E_{1-2s}(\tau)= \frac{1}{2\zeta(2s)} \cE_s(\tau)\ ,\qquad
%\cE_s = \sum_{(m,n)\neq (0,0)}
%\left(
%\frac{\tau_2}{|m+n\tau|^2}
%\right)^s \ ,
%\ee
and satisfies the functional relation
\be
\xi(s)\, \cE_s(\tau) = \xi(-s) \cE_{-s}(\tau)\ .
\ee
The minimal Eisenstein series \eqref{eismin} can be obtained 
from the principal Eisenstein series \eqref{eisgen} by taking the limit 
\be
E(g;s) = 2 \zeta(2s) \lim_{s_2\to 0}  E(g;s_1=s,s_2)
\ee
or more generally by keeping the leading term 
in the limit $\lambda_{ij}\to 1$ for any $i\neq j$.
Thus, its constant terms are given by
\be
\begin{split}
\frac{E_{P_{\rm min}}(g;s)}{2\zeta(2s)}  &= \left( \frac{t_1 t_2}{t_3^2}\right)^{\frac{2s}{3}}
+\frac{\xi(2s-1)}{\xi(1-2s)} \frac{t_2}{t_3}  \left( \frac{t_1 t_3}{t_2^2}\right)^{\frac{2s}{3}}
+\frac{\xi(2s-2)\xi(2s-1)}{\xi(1-2s)\xi(2-2s)} \frac{t_1 t_2}{t_3^2}
\left( \frac{t_1 t_3}{t_2^2}\right)^{\frac{2s}{3}}\ ,\\
\frac{E_{P_{\rm max}}(g;s)}{2\zeta(2s)}  &= \frac{t_1^s E_s(\tau)}{2\zeta(2s)} 
+ t_1^{3-2s} \frac{\xi(2s-2)\xi(2s-1)}{\xi(1-2s)\xi(2-2s)} \ ,
\end{split}
\ee
where in the second line we used the identity $E_0(\tau)=-1$ \cite{Obers:1999um}. 

It is straightforward if tedious to check that all terms in these expansions have the
same values \eqref{casgen} of  the quadratic and cubic Casimirs, where $\cC_2$
and $\cC_3$ are given by 
\be
\label{c2c30}
\cC_2=-\frac12(E_{p}\,F_{p}+F_{p}\,E_{p}+\,E_{q}\,F_{q}+\,F_{q}\,E_{q}+ F\,E+ E\,F)
-\frac13( H_{p}^2 + H_{p}\,H_{q}+H_{q}^2)\\
\ee
\be
\label{c2c3}
\begin{split}
\cC_3&=\frac1{54}(3\,E_{p}\,F_{p}\,H_{p}+6\,E_{p}\,F_{p}\,H_{q}-9\,E_{p}\,E_{q}\,F+3\,E_{p}\,H_{p}\,F_{p}+6\,E_{p}\,H_{q}\,F_{p}-9\,E_{p}\,F\,E_{q}\\
&+3\,F_{p}\,E_{p}\,H_{p}+6\,F_{p}\,E_{p}\,H_{q}+9\,F_{p}\,F_{q}\,E+3\,F_{p}\,H_{p}\,E_{p}+6\,F_{p}\,H_{q}\,E_{p}+9\,F_{p}\,E\,F_{q}\\
&-9\,E_{q}\,E_{p}\,F-6\,E_{q}\,F_{q}\,H_{p}-3\,E_{q}\,F_{q}\,H_{q}-6\,E_{q}\,H_{p}\,F_{q}
-3\,E_{q}\,H_{q}\,F_{q}-9\,E_{q}\,F\,E_{p}\\
&+9\,F_{q}\,F_{p} \,E-6\,F_{q}\,E_{q}\,H_{p}-3\,F_{q}\,E_{q}\,H_{q}-6\,F_{q}\,H_{p}\,E_{q}-3\,F_{q}\,H_{q}\,E_{q}+9\,F_{q}\,E\,F_{p}\\
&+3\,H_{p}\,E_{p}\,F_{p}
+3\,H_{p}\,F_{p}\,E_{p}-6\,H_{p}\,E_{q}\,F_{q}-6\,H_{p}\,F_{q}\,E_{q}-4\,H_{p}\,H_{p}\,H_{p}-2\,H_{p}\,H_{p}\,H_{q}\\
&-2\,H_{p}\,H_{q}\,H_{p}+2\,H_{p}\,H_{q}\,H_{q}
+3\,H_{p}\,F\,E+3\,H_{p}\,E\,F+6\,H_{q}\,E_{p}\,F_{p}+6\,H_{q}\,F_{p}\,E_{p}\\
&-3\,H_{q}\,E_{q}\,F_{q}-3\,H_{q}\,F_{q}\,E_{q}-2\,H_{q}\,H_{p}\,H_{p}+
2\,H_{q}\,H_{p}\,H_{q}+2\,H_{q}\,H_{q}\,H_{p}
+4\,H_{q}\,H_{q}\,H_{q}\\
&-3\,H_{q}\,F\,E-3\,H_{q}\,E\,F-9\,F\,E_{p}\,E_{q}-9\,F\,E_{q}\,E_{p}
+3\,F\,H_{p}\,E-3\,F\,H_{q}\,E+3\,F\,E\,H_{p}\\
&-3\,F\,E\,H_{q}+9\,E\,F_{p}\,F_{q}+9\,E\,F_{q}\,F_{p}+3\,E\,
H_{p}\,F-3\,E\,H_{q}\,F+3\,E\,F\,H_{p}-3\,E\,F\,H_{q}).
\end{split}
\ee

\section{Fourier expansion of the minimal Eisenstein series}
\label{App:MinimalFourier}
 
As an illustration of the general principle explained in Section \ref{secnonab}, 
let us compute the non-Abelian Fourier expansion of the minimal Eisenstein
series \eqref{eismin}:
\be
\label{esl3}
E(g,s)= %\sum_{(m_1,m_2,m_3)\in\IZ^3}
\sum'
\left[ t\, e^{-\phi}  \left(m_1 +  \zeta m_2 -\frac12 (\sigma-\zeta\tzeta) m_3\right)^2
+\frac{1}{t^{2}} \left(m_2 + \tzeta m_3\right)^2 + t\, e^\phi  m_3^2 \right]^{-s}
\ee
where the sum runs over $(m_1,m_2,m_3)\in\IZ^3\backslash(0,0,0)$. 
We first split
\be
\label{spl1}
E(g,s)=E^{(0)} + E^{(1)} 
\ee
where the first term is the contribution with $m_3=0$ (and therefore
$(m_1,m_2)\in\IZ^2\backslash(0,0)$) and the second term is the one
with $m_3\neq 0$. The first term is proportional to the standard $SL(2,\IZ)$ Eisenstein 
series \eqref{sl2eis}, 
\be
\label{dsl2}
\begin{split}
E^{(0)}% &= \nu^{-s/3} \sum'_{m_1,m_2} \left[ \frac{| m_1 +  \tau m_2 |^2}{\tau_2}  \right]^{-s} \ \\
&= \nu^{-s/3} \left[ 2\zeta(2s)\tau_2^s + 2\sqrt{\pi} \tau_2^{1-s}
\frac{\Gamma(s-1/2)}{\Gamma(s)} \zeta(2s-1) \right. \\
&\left. \qquad\qquad +  \frac{2 \pi ^s \tau_2^s}{(2\pi)^{1/2-s}\Gamma(s)}
\sum_{\tilde m_1\ne 0} \sum_{m_2\ne 0}
\left| \tilde m_1 \right|^{2s-1}
\cK_{s-1/2} \left(2\pi \tau_2 |\tilde m_1 m_2|\right)
e^{2\pi i \tilde m_1  m_2 \tau_1} \right]\ .
\end{split}
\ee
The last term in the bracket corresponds to the Abelian Fourier coefficient with 
$(p,q)=(0,\tilde m_1 m_2)$, corresponding to D(-1)-instantons, while the first
two terms reproduce the first two constant terms on the first line of \eqref{minpmax}.

In the second term of \eqref{spl1}, the sum over $(m_1,m_2)$ runs over $\IZ^2$ without restriction:
thus we may perform a Poisson resummation over $(m_1,m_2)$ by  
using the standard integral representation of the summand (see e.g. \cite{Obers:1999um}),
\be
M^{-s}=\f{\pi^s}{\Gamma(s)}\int_0^\infty \frac{du}{u^{s+1}}e^{-\frac{\pi}{u} M} .
\ee
After Poisson resummation we then obtain 
\be
\begin{split}
E^{(1)} &=  \frac{\pi^s \sqrt{t}\, e^{\phi/2}}{\Gamma(s)}\
\sum_{\tilde m_1,\tilde m_2} \sum'_{m_3}
\int_0^\infty  \frac{du}{u^{s}} 
\exp\left[ -\frac{\pi u}{t} e^{\phi}  \tilde m_1^2 -\pi u\, t^2
\left(\tilde m_2 -\zeta \tilde m_1\right)^2\right. \\
&\qquad \qquad \left.  -\frac{\pi\, t}{u}  e^\phi m_3^2
-2\pi\I \left(  \tzeta \tilde m_2 m_3 -\frac12 \tilde m_1 m_3 (\sigma+\zeta\tzeta)\right) \right].
\end{split}
\ee
The term with $(\tilde m_1,\tilde m_2)=(0,0)$ leads to a Gamma-type integral, 
\be
E^{(1)}  = 2\pi \left ( t \,e^\phi\right)^{\frac32-s} \frac{\Gamma(s-1)\zeta(2s-2)}{\Gamma(s)}
+ E^{(2)} 
\ee
while the one
with $(\tilde m_1,\tilde m_2)\neq (0,0)$ leads to a Bessel function:
\be
\begin{split}
E^{(2)}  &=  \frac{2 \pi^s (t\, e^\phi)^{\frac32-s}}{(2\pi)^{s-1}\Gamma(s)}
\sum'_{\tilde m_1,\tilde m_2} \sum'_{m_3}
|m_3|^{2(1-s)}
 e^{-2\pi\I  \tzeta \tilde m_2 m_3 +\I \pi  \tilde m_1 m_3 (\sigma+\zeta\tzeta) } \\
 & \qquad\qquad \cK_{1-s}\left( 2\pi e^{\phi/2} | m_3| 
\sqrt{ e^\phi\, \tilde m_1^2 + t^3 \left(\tilde m_2 - \zeta \tilde m_1\right)^2}
\right)
\end{split}
\ee
or,  in a manifestly $SL(2,\IZ)$-invariant form,
\be
%\begin{split}
E^{(2)}  =  \frac{2 \pi^s \, \nu^{-1+\frac23s}}{(2\pi)^{s-1}\Gamma(s)}
\sum'_{\tilde m_1,\tilde m_2} \sum'_{m_3}
|m_3|^{2(1-s)}
\, e^{-2\pi\I  m_3 ( \tilde m_1 \psi +   \tilde m_2 c_0) }  
%\\ & 
 \quad \cK_{1-s}\left( 2\pi  \frac{| m_3|}{ \sqrt{\nu}}\cdot
\frac{| \tilde m_2 - \tau \tilde m_1|}{\sqrt{\tau_2}  }
\right).
%\end{split}
\ee
The term with $\tilde m_1=0$ is an Abelian Fourier coefficient with 
$(p,q)=(\tilde m_2 m_3,0)$, corresponding to D5-brane instantons:
\be
\label{d5i}
E^{(3)}  =  \frac{2 \pi^s (t\, e^\phi)^{\frac32-s}}{(2\pi)^{s-1}\Gamma(s)}
\sum'_{\tilde m_2} \sum'_{m_3}
|m_3|^{2(1-s)}
 e^{-2\pi\I  \tzeta \tilde m_2 m_3}
\, \cK_{1-s}\left( 2\pi e^{\phi/2} t^{3/2} | \tilde m_2 m_3|  \right)\ .
\ee
The general term with  $\tilde m_1\neq 0$ can be recast as \eqref{nonab2}, by identifying
\be
\tilde m_1 m_3 = - k\ ,\qquad 
\tilde m_2 m_3 = - k m\ ,\qquad m\in \IZ + \frac{\ell'}{|k|}\ ,
\ee
%and
\be
\label{naco1}
\tilde\Psi_{k,\ell'} (\zeta) = \left( \sum_{d| \, |k|} d^{2(1-s)} \right)
\frac{2\pi^s \, (t\, e^{\phi})^{\frac32-s} }{(2\pi)^{s-1}\Gamma(s)}\, 
\cK_{1-s}\left( 2\pi  |k| e^{\phi/2} \sqrt{ e^\phi  + \zeta ^2 t^3 } \right) \ .
\ee

\subsection*{Dual expansion}

Alternatively, one may arrive at the non-Abelian Fourier expansion \eqref{nonab}
by returning to \eqref{esl3}, extracting the term with $m_2=m_3=0$ and performing
a Poisson resummation over the single variable $m_1$:
\be
\begin{split}
E(g,s) &= 2 \zeta(2s) \, e^{s\phi}\, t^{-s} + \sum_{\tilde m_1} \sum'_{m_2,m_3}
\frac{e^{\phi/2}}{\sqrt{t}} \frac{\pi^s}{\Gamma(s)}\int_0^\infty  \frac{du}{u^{s+\frac12}} 
\exp\left[ -\frac{\pi u}{t} e^{\phi}  \tilde m_1^2 \right. \\
&\left. -\frac{\pi}{u t^2}
\left(m_2 + \tzeta m_3\right)^2 -\frac{\pi\, t}{u}  e^\phi m_3^2
+2\pi\I \tilde m_1 \left(  \zeta m_2 -\frac12 (\sigma-\zeta\tzeta) m_3\right) \right].
\end{split}
\ee
For $\tilde m_1=0$, one may similarly extract the term with $m_3=0$, Poisson resum
over $m_2$ and extract the term $\tilde m_2=0$, to get
\be
\begin{split}
\frac{2 \pi^s \, \Gamma(s-\frac12)\, e^{\phi/2} \,t^{2s-\frac32} }{\pi^{s-\frac12}\Gamma(s) }
\zeta(2s-1) + 
\frac{2\,\pi^s \Gamma(s-1) (t\,e^{\phi})^{\frac32-s} }{\pi^{s-1}\Gamma(s) }
\zeta(2s-2)\\
+ \frac{2\, \pi^s(e^{\phi} t)^{\frac32-s}}{(2\pi)^{s-1}\Gamma(s)} 
\sum'_{\tilde m_2} \sum'_{m_3} |m_3|^{2(1-s)} \cK_{1-s}\left( 2\pi e^{\phi/2} t^{3/2}
|\tilde m_2 m_3| \right) \,e^{-2\pi \I  \tzeta \tilde m_2 m_3} .
\end{split}
\ee
The last term is the Abelian Fourier coefficient with $(p,q)=(\tilde m_2 m_3,0)$,
identical to \eqref{d5i}, 
corresponding to D5-instantons.
For $\tilde m_1\neq 0$, the integral over $u$ is of Bessel type, leading to
\be
\begin{split}
\frac{2\, \pi^s\, e^{s \phi}}{t^s\, (2\pi)^{1/2-s}\Gamma(s)}&
\sum'_{\tilde m_1} \sum'_{m_2,m_3} |\tilde m_1|^{2s-1}
e^{2\pi\I \tilde m_1 \left( \zeta m_2 -\frac12 (\sigma-\zeta\tzeta) m_3\right) } \\
 & \qquad\qquad \cK_{s-1/2}\left( \frac{2\pi e^{\phi/2}}{t^{3/2}} |\tilde m_1| 
\sqrt{ \left(m_2 + \tzeta m_3\right)^2 + t^3\, e^\phi \, m_3^2 }
\right) .
\end{split}
\ee
For $m_3=0$, this reduces to the Abelian Fourier coefficient $(p,q)=(0,\tilde m_1 m_2)$,
corresponding to D(-1)-instantons, identical to the last term in \eqref{dsl2},
\be
\frac{2\, \pi^s\, e^{s \phi}}{t^s\, (2\pi)^{1/2-s} \Gamma(s)}
\sum'_{\tilde m_1} \sum'_{m_2} |\tilde m_1|^{2s-1}
\, \cK_{s-1/2}\left( \frac{2\pi e^{\phi/2}}{t^{3/2}} |\tilde m_1 m_2| 
\right) \, e^{2\pi\I  \tilde m_1  m_2 \zeta} .
\ee
For the general term with $\tilde m_1\neq 0$ and $m_3\neq 0$, identifying 
\be
\tilde m_1 m_3 = k\ ,\qquad 
\tilde m_1 m_2 = - k n \ ,\qquad n\in \IZ + \frac{l}{|k|} \ ,
\ee
we recognize the non-Abelian Fourier coefficient \eqref{nonab} with 
\be
\label{naco2}
\Psi_{k,\ell} (\tzeta) = \left( \sum_{d| \, |k|} d^{2s-1} \right)
 \frac{2\, \pi^s e^{s \phi} }{(2\pi)^{1/2-s} \Gamma(s)\, t^s }\,
\cK_{s-1/2}\left( 2\pi   |k|  e^{\phi/2} \sqrt{  e^\phi  +  t^{-3}\, \tzeta ^2 } \right) \ .
\ee
Thus, we have reproduced the non-Abelian Fourier coefficients summarized in 
\eqref{psimin}.

%Inserting this into our previous result for $\C_{y, x_0, x_1}^{s_1, s_2, s_3}$ in Eq. (\ref{nonAbeliancoefficientSP}) we find that the Abelian Fourier coefficients of Proskurin become
%\be
%c^{s_1, s_2, s_3}_{x_0, x_1}\sim \f{d^{2(1-s_1)} x_1^{4s_1+2s_2-5}}{(x_0^2+x_1^2)^{1/4}} \exp\left[- \f{2\pi(x_0^2+x_1^2)^{3/2}}{x_0^2}\right].
%\ee

\section{Asymptotic expansion of the integral $\mc{I}$\label{SaddlePoint}}

In this appendix, we discuss the properties of an integral which is
a key ingredient for the Fourier expansion, as it enters
the coefficients (\ref{genab}) and (\ref{nonAbeliancoefficient}): 
\be
\label{defI}
 \mc{I}_{s_1,s_2}(R,\x)
 =  \int_0^{\infty} K_{s_3-1/2}\big(2\pi R\,  \x^{-1}  \sqrt{1+x}\big) \,K_{s_3-1/2}\Big(2\pi R \, \x^{1/2} \sqrt{1+1/x}\Big) \, x^{\f{s_2-s_1}{2}} \f{dx}{x}\ .
 \ee
 First, we note the functional equation
 \be
 \mc{I}_{s_1,s_2}(R,\x)=\mc{I}_{s_2,s_1}\left(\frac{R}{\sqrt{\x}}, \frac{1}{\x}\right)\ ,
 \ee
 which, at the origin of moduli space $\tau_2=\nu=1$, amounts to exchanging $(p,q)$
 in \eqref{pqvariables}.

In order to find the semi-classical interpretation of the Fourier coefficients, we shall be 
interested in the limit $R\to\infty$ keeping $\x$ fixed. In this regime, the integral 
\eqref{defI} can be evaluated in the saddle point approximation. For large argument, 
the Bessel function may be approximated by 
 \be
 K_s(x)\sim \sqrt{\f{\pi}{2x}}e^{-x} \left( 1 +\frac{4s^2-1}{8x} + \mathcal{O}(1/x^2) \right) \ .
 \label{BesselExpansion}
 \ee
 To leading order, the integral then simplifies to
 \be
 \mc{I}_{s_1,s_2}(R,\x)\sim \frac{\x^{1/4}}{4R}
  \int_0^{\infty}\f{ x^{\f{s_2-s_1}{2}-\frac34}\ e^{-2\pi S(x)}}{\sqrt{1+x}}  dx\ ,
 \ee
 where $S(x)$ is given by
 \be
S(x) = R \left(  {\bf x}^{-1} \sqrt{1+x} +  {\bf x}^{1/2} \sqrt{1+\frac{1}{x}} \right)\ .
 \ee
 The exponent is extremized at $x={\bf x}$, with 
 \be 
 S(\x)=\frac{ R(1+\x)^{3/2}}{\x}\ ,\qquad \pa^2_x S (\x) = \frac{3 R}{4\x^2 \sqrt{1+\x}}\ .
 \ee
The saddle point approximation  is then
\be
\begin{split}
\label{sadap}
\mc{I}_{s_1,s_2}(R,\x) &\sim 
\f{\x^{\f{s_2-s_1-1}{2}}  }{4R\sqrt{1+\x}} \left[
\frac12 \pa^2_x S(\x) \right]^{-1/2}   \exp\left(-2\pi S(\x) \right)
\left( 1 + \frac{I_1}{R}+ \mathcal{O}(1/R^2) \right)\\
 &\sim 
\f{\x^{\f{s_2-s_1+1}{2}}  }{\sqrt6\, R^{3/2} (1+\x)^{1/4}}  \exp\left(- \frac{ 2\pi R(1+\x)^{3/2}}{\x}\right)
\left( 1 + \frac{I_1}{R}+ \mathcal{O}(1/R^2) \right).
\end{split}
\ee
The subleading term $I_1$ can be computed by exponentiating
the prefactor\footnote{Note that the subleading term in the Bessel function does not
contribute at this order.},
\be
\tilde S(x) = S(x) - \frac{1}{2\pi} \log\left[ x^{\f{s_2-s_1}{2}-\frac34}/\sqrt{1+x}\right] 
\ee
and expanding around the perturbed saddle point,
\be
x = \x - \frac{\x}{6\pi R \sqrt{1+\x}} \left( 3+5\x+2(1+\x)(s_1-s_2) \right) + \frac{1}{R^2} {\delta x}\ .
\ee
We find 
\be
\begin{split}
\tilde S(x) &= \tilde S(\x) +  
\frac{3 {\delta x}^2}{8 \x^2 \sqrt{\x+1}}-\frac{{\delta x}^3
   (7 \x+5)}{16 \left(\x^3
   (\x+1)^{3/2} \sqrt{R} \right)} +\frac{{\delta x}^2}{128 \pi R \x^4(\x+1)^{5/2}} \times \\
   &\times\left(\pi  {\delta x}^2
   \left(59 \x^2+85 \x+35\right)+4 \x^2
   (\x+1)^{3/2} (2 {s_1} (5 \x+3)-2 {s_2} (5
   \x+3)+25 \x+9)\right)+ \dots
        %+\mathcal{O}(1/R^{3/2})
   \end{split} 
\ee
Expanding the non-Gaussian piece and performing the Gaussian integration term by term,
we find that the leading quantum correction is given by
\be
\begin{split}
I_1&=\frac{2048 \x^8 (\x+1)^{7/2}}{2187 \pi    R}  \left(-\frac{81 (2 {s_1}
   (5 \x+3)-2 {s_2} (5 \x+3)+25 \x+9)}{128
   \x^4 (\x+1)^3} \right. \\
   &\qquad \left. -\frac{27 \left(59 \x^2+85
   \x+35\right)}{16 \x^2 (\x+1)^{7/2}}+\frac{15 (7
   \x+5)^2}{(\x+1)^3}\right) \ .
\end{split}
\ee

\section{Non-Abelian Fourier expansions and representation theory\label{ap_rep}}

In this appendix we take a representation theoretic viewpoint on the non-Abelian
Fourier expansions discussed in Section \ref{secnonab}. The starting point is that,
on the non-Abelian Fourier expansion \eqref{nonab}, the Heisenberg algebra acts
as
\be
E_p= - 2\pi \I k n \ ,\qquad E_q= -\pa_n\ , \qquad E= - 2\pi \I k\ ,
\ee
while on the dual Fourier expansion \eqref{nonab2},
\be
E_p= \pa_m\ ,\qquad E_q= 2\pi \I k m\ \ , \qquad E= 2\pi \I k\ .
\ee
More generally, when $\Psi(t,\phi;\zeta,\tzeta,\sigma)$ is  an automorphic form for $SL(3,\IZ)$, 
the non-Abelian expansions  \eqref{nonab} and \eqref{nonab2} correspond to two different
choices of polarization in writing $\Psi$ as a matrix element \eqref{overl}, where either
$(E_p,E)$ or $(E_q,E)$ have been diagonalized. Note that with the exception of the 
minimal representation, $(E_p,E)$ do not form a complete basis of commuting operators,
which is responsible for the appearance of additional quantum numbers such as $q$ in
\eqref{nonAbeliancoefficient}.

\subsection{Minimal Eisenstein series \label{ap_min}}

Using this observation, it is easy to see that the non-Abelian Fourier expansion of the minimal
Eisenstein series \eqref{eismin} in the polarization  \eqref{nonab2} can be written as 
an inner product \eqref{overl} of a $G(\IZ)$ invariant vector $f_\IZ$ 
with the transformed spherical vector
$\rho(g^{-1}) f_K$, where $\rho$ acts on functions of two variables $y=2\pi k$ and $x_0=2\pi k m$
via 
\be
\begin{array}{lcll}
E_{p}=y \p_{x_0}\, ,&\qquad&
F_{p}=&-x_0 \p_y \, ,\\[2mm]
E_{q}=\I x_0\, ,&&
F_{q}=&-\I(x_0\p_{x_0}+y\p_y+(3-2s))\p_{x_0}\\[2mm]
E=\I y\, ,&&
F=&-\I (x_0\p_{x_0}+y\p_y+(3-2s))\p_y\\[2mm]
H_p=x_0\p_{x_0}-y\p_y\, ,&&H_q=&-2x_0\p_{x_0}-y\p_y-(3-2s)\, ,
\end{array}
\label{sl3min1}
\ee
and the spherical vector $f_K$ and the $G_\IZ$ invariant
distribution are given by \cite{Pioline:2003bk}
\be
 f_K
={\cal K}_{1-s}\Big(\sqrt{y^2+x_0^2}\Big)\, ,\qquad
f_\IZ
=\mu_{2-2s}(y,x_0)\ .
\label{fsph1}
\ee

Similarly, the non-Abelian Fourier expansion in the polarization \eqref{nonab} can be written as 
\eqref{overl} where $\rho$ is now the representation on functions of two
variables $y=-2\pi k$ and $x_0=-2\pi k n$, 
\be
\begin{array}{lcll}
E_{p}=\I x_0\, ,&&
F_{p}=&-\I (x_0\p_{x_0}+y\p_y+2s)\p_{x_0}\\[2mm]
E_{q}=-y \p_{x_0}\, ,&\qquad&
F_{q}=&x_0 \p_y \, ,\\[2mm]
E=\I y\, ,&&
F=&-\I (x_0\p_{x_0}+y\p_y+2s)\p_y\\[2mm]
H_q=x_0\p_{x_0}-y\p_y\, ,&&H_p=&-2x_0\p_{x_0}-y\p_y-2s\, ,
\end{array}
\label{sl3min2}
\ee
obtained from the previous one by Fourier transform over $y$ and $x_0$.
The spherical vector is now
\be
f_K
={\cal K}_{s-\frac12}\Big(\sqrt{y^2+x_0^2}\Big)\ , \qquad f_\IZ
=\mu_{2s-1}(y,x_0)\ ,
\label{fsph2}
\ee
in agreement with $\Psi_{k,\ell}$ in \eqref{psimin}.
The spherical vector $f_K$ in a representation where the Heisenberg algebra takes
the canonical form \eqref{sw1} or \eqref{sw2} is sometimes called the ``generalized
Whittaker vector'' in the mathematics literature.

\subsection{Principal Eisenstein series \label{ap_prin}}

The generalized Eisenstein series \eqref{eisgen}
is attached to the general continuous representation 
\bea\nn
\begin{array}{lr}
E_{q}=-\partial_x +w \partial_v \ , &
F_{q}=-x^2\partial_x-v\partial_w+2 s_1 \,x \, ,\\[1mm]
E_{p}=\partial_w \ , \qquad&\hspace{-1cm}
F_{p}=w^2\partial_w+vw\partial_v-(v+xw)\partial_x
+2s_2\, w\, ,\\[1mm]
E=\partial_v \ , &\hspace{-5.0cm}
F=v^2\partial_v+vw\partial_w+x(v+xw)\partial_x 
+2(s_1+s_2)v-(1-2s_1)xw\, ,\\[1mm]
H_q=2x\partial_x+v\partial_v-w\partial_w+2s_1\, , \qquad&
H_p=-x\partial_x+v\partial_v+2w\partial_w+2s_2\, 
\end{array}
\label{sl3cts}
\eea
with spherical vector
\be
f_K = \left[1+x^2+(v+xw)^2\right]^{-s_1} 
\ \left[1+v^2+w^2\right]^{-s_2}\ .
\label{spherical_A2}
\ee
Equivalently,  it can be written in such a way that the Heisenberg generators
are represented as in \eqref{sw1}, and  the $SL(2,\IZ)$ action
acts linearly on $(y,x_0)$:
\be
\begin{split}
E_{p}&=y \p_{x_0}\, ,\qquad E_{q}=\I x_0\, ,\qquad E=\I y\, ,\qquad
F_{p}=-x_0 \p_y + \I  \frac{x_1^3}{y^2}\, ,\\
H_p&=x_0\p_{x_0}-y\p_y\ ,\quad
H_q=-2x_0\p_{x_0}-y\p_y-x_1 \p_{x_1}+2(2s_1+s_2-3)\, ,\\
F_{q}&=-\I(x_0\p_{x_0}+y\p_y+x_1 \p_{x_1} +2)\p_{x_0}
+\left(4-4s_1-2s_2\right) \left( \frac{y}{9x_1}  \pa_{x_1}^2 - \I \pa_0\right)\\
&\qquad\qquad+\frac{2}{27} (3s_1-2)(6s_1+6s_2-7)\frac{y}{x_1^2} \pa_{x_1} +\frac{y}{27}\pa_{x^1}^3
\label{sl3x1gen}
\end{split}
\ee
and $F=[F_p,F_q]$. For $(s_1,s_2)=0$, this reproduces the representation 
obtained in \cite{Pioline:2004xq} by restricting the minimal representation of 
$E_6$ constructed in \cite{Kazhdan:2001nx} to singlets of the first two
factors in the maximal subgroup $SL(3)\times SL(3)\times SL(3)\subset E_6$. 
For $(s_1,s_2)=(2/3,2/3)$ one recovers instead the minimal representation
of $G_2$ considered in \cite{MR0342049,Gunaydin:2007qq}. 
%For $s_1=0$, the sequence of transformations relating \eqref{sl3x1gen} to
%\eqref{sl3cts} was given in \cite{Pioline:2003bk,Pioline:2004xq}.
%Although this sequence breaks down in the case $s_1\neq 0$ relevant for this paper,
%\eqref{sl3x1gen} and the general continuous representation have the same infinitesimal
%character and it should be possible in principle to find the intertwiner between them.

We can now frame the non-Abelian Fourier expansion in the general framework \eqref{overl},
and determine the real spherical vector for the principal series away from the semi-classical limit. 
For this purpose, we  change of variables to
 \be
 y=-k\ ,\qquad x_0=p\ ,\qquad
 x_1=(d^2 q)^{1/3}\ , 
 \label{ChangeOfVariables}
 \ee
and work at the origin of moduli space
where $\tau_1=0, \tau_2=\nu=1$, such that
\be
[\tau_2]_{-k,p}=\f{d^2}{y^2+x_0^2}\ , \qquad R_{d,q}=x_1\ ,\qquad  \x_{d,q}=\f{y^2+x_0^2}{x_1^2}\ .
\ee
Moreover, the phase factor in the non-Abelian term of \eqref{nonAbeliancoefficient} becomes
\be
e^{-2\pi \I q [\tau_1]_{-k,p}}=e^{\f{2\pi \I q d \al}{k}}e^{\f{2\pi \I\ pq}{k(k^2+p^2)}}
= e^{\f{2\pi \I x_1^3 \al }{d k}} \, e^{ -\f{2\pi \I x_0 x_1^3}{y(y^2+x_0^2)}}.
\label{decomposition}
\ee
We can therefore write the non-Abelian part of the expansion at the origin of moduli space as
the overlap 
\be
E_{NA}(1; s_1, s_2)= \sum_{(y,x_0,x_1^{3})\in \mbb{Z}^*\times \mbb{Z}\times \mbb{Z}}
 f_{\IZ}(y, x_0, x_1) \, f_K(y, x_0, x_1) + \dots
\label{NonAbelianTermSPApprox}
\ee
where the real spherical vector is given by
\be
\label{exactsl3}
\begin{split}
f_K(y,x_0,x_1) &= \, 
(y^2+x_0^2)^{\frac12(s_1-s_2-1)} \, x_1^{3(s_1+s_2-1)} \, e^{-\f{2\pi \I x_0 x_1^3}{y(y^2+x_0^2)}}\\
&\times \int_0^{\infty} K_{s_3-\frac12}\Big(\frac{2\pi \,x_1^3}{y^2+x_0^2}\sqrt{1+x}\Big) 
K_{s_3-\frac12}\Big(2\pi \sqrt{(y^2+x_0^2)(1+\frac{1}{x})}\Big) x^{\f{s_2-s_1}{2}} \f{dx}{x} \ ,
 \end{split}
 \ee
the summation  measure (or ``adelic spherical vector")  is
\be
f_{\IZ}(y, x_0, x_1)\equiv\f{4 e^{\f{2\pi \I x_1^3 \al }{d k}}}{\sqrt{6}} 
\f{|d|^{\f{5}{6}-2s_1}|x_1|^{4s_1+2s_2-\f{5}{2}} }{\xi(2s_1)\xi(2s_2)\xi(2s_3)} 
\sum_{d_1|d} \sum_{d_2|\frac{d}{d_1}}  d_1^{1-2s_3} d_2^{1-2s_2}\sigma_{1-2s_1,1-2s_3}\Big(\f{d}{d_1 d_2}, \f{x_1^3}{d^{2}}\Big)\ ,
\label{adelicsph}
\ee
and the ellipses stand for degenerate contributions with support at $y=0$. 
In \eqref{adelicsph}, we recall that $d\equiv \gcd(y, x_0)$, and that $f_{\IZ}(y, x_0, x_1)$
vanishes unless $d^2$ divides $x_1^3$. We note that the real and $p$-adic spherical
vectors for the principal series of $SL(n,\IR)$ for any $n$ have been obtained in
\cite{MR0407208,MR581582,MR1767115,gerasimov-2008}. It would be interesting to see how 
\eqref{adelicsph} emerges as a product of the  $p$-adic spherical
vectors over all primes.

The spherical vector simplifies considerably in the limit where $y,x_0,x_1$ are scaled to infinity
with fixed ratio: in this case the saddle point approximation \eqref{sadap} becomes
\be
\mc{I}(y, x_0, x_1) \sim  \f{(y^2+x_0^2)^{\f{s_2-s_1+1}{2}} x_1^{s_1-s_2-2}}{(y^2+x_0^2+x_1^2)^{1/4}}\exp\left[- \f{2\pi (y^2+x_0^2+x_1^2)^{3/2}}{y^2+x_0^2}\right]\ ,
\ee
and  the spherical vector simplifies to 
\be
f_K(y, x_0, x_1)\sim  
 \f{x_1^{4s_1+2s_2-5}}{(y^2+x_0^2+x_1^2)^{1/4}} \exp\left[- \f{2\pi(y^2+x_0^2+x_1^2)^{3/2}}{y^2+x_0^2}
-\f{2\pi \I x_0 x_1^3}{y(y^2+x_0^2)}\right]\ .
\label{nonAbeliancoefficientSP}
\ee
As a consistency check, we note that in the special case $(s_1, s_2)=(0,0)$ (i.e. 
 $(\lambda_{23}, \lambda_{21})=(1,1)$), this result agrees with the semi-classical 
 spherical vector of the principal series representation of $SL(3,\IR)$ obtained by 
restricting the minimal representation of $E_6$ singlets of the first two
factors in the maximal subgroup $SL(3)\times SL(3)\times SL(3)\subset E_6$. \cite{Pioline:2004xq}

Moreover, we note that \eqref{nonAbeliancoefficientSP} is in fact a special case of the
general formula for the spherical vector (or lowest K-type) of the minimal representation 
of any group $G$ viewed as a quasiconformal
group $G={\rm QConf}(J)$  \cite{Kazhdan:2001nx},
\be
\begin{split}
f_{K}
&\sim  \exp\left[ -
\sqrt{y^2+x_0^2} \left( 
1 + 3 \frac{ (x^a)^2}{y^2+x_0^2}
+ \frac1{12} \frac{ (\kappa_{abc} x^b x^c)^2}{(y^2+x_0^2)^2}
+ \frac1{2^2\cdot 3^2} \frac{ (\kappa_{abc} x^a x^b x^c)^2}{(y^2+x_0^2)^3}
\right)^{1/2}\right.\\
&\left. \qquad\qquad + \I \frac{x_0} {6y (y^2+x_0^2)} \kappa_{abc}x^a x^b x^c\right]\ .
\label{sphming}
\end{split}
\ee
Indeed, \eqref{sphming} reduces to \eqref{nonAbeliancoefficientSP} in the one-modulus case
with $\kappa_{111}=6$, corresponding to $G=G_{2(2)}$. This is in accord with the fact that
the minimal representation of $G_{2(2)}$ is an irreducible representation of $SL(3,\IR)
\subset G_{2(2)}$ in the non-spherical supplementary series (see discussion
in Section 3.6). Note that the exact lowest K-type of the minimal representation of $G_2$ was found 
in \cite{Gunaydin:2007qq}, Eq. (3.119): it would be interesting to see if the
integral in \eqref{exactsl3} can be similarly evaluated in closed form.
Moreover, the exact spherical vector of the minimal representation of any simply-laced
group $G$ in its split form was found in  \cite{Kazhdan:2001nx}. It would be interesting
to see what representation of $SL(3,\IZ)$ is obtained in the $G_5={\rm Str_0(J)}$ invariant
sector, and see how \eqref{exactsl3} is reproduced.

We conclude with a comment on the ``Abelian limit'' $y \rightarrow 0$ , which is 
needed to properly extract the Abelian Fourier coefficients $\Psi_{p,q}$ in (\ref{genab}).
As already discussed in \eqref{vtransformedcomb}, the phase factor 
in $f_K$ is singular, but so is the measure $f_\IZ$, and the two singularities
cancel. Thus, we may define $\tilde f(0,x_0, x_1)$ as the $y\to 0$  limit 
of  $f(y,x_0, x_1)$ after removing the singular phase  \cite{MR2094111},
\be
\tilde f(0,x_0, x_1)\equiv \lim_{y\rightarrow 0}\left( \exp\left[ \f{2\pi \I  x_0 x_1^3}{y(y^2+x_0^2)}\right] f(y, x_0, x_1)\right) \ ,
\label{limit}
\ee
and perform the  opposite operation for the dual vector.
In particular, the spherical vector \eqref{exactsl3} reduces in this limit to
\be
\begin{split}
\tilde f_K(0,x_0, x_1)
&=x_0^{s_1-s_2-1} \, x_1^{3(s_1+s_2-1)} \\
&\times \int_0^{\infty} K_{s_3-\frac12}\Big(\frac{2\pi \,x_1^3}{x_0^2}\sqrt{1+x}\Big) 
K_{s_3-\frac12}\Big(2\pi x_0 \sqrt{1+1/x}\Big) x^{\f{s_2-s_1}{2}} \f{dx}{x} \ .
\end{split}
\label{limitfk}
\ee
%SHOULD WE COMPARE TO GIVENTAL, EQ 6.33 OF SL3 NOTES ?

%\bibliography{../common/combined}
%\bibliographystyle{../common/utphys}

\providecommand{\href}[2]{#2}\begingroup\raggedright\endgroup

\end{document}